\documentclass[twocolumn,floatfix,pra,aps,showpacs]{revtex4-1}
\usepackage{graphicx,amsmath,amssymb,xcolor}
\usepackage{nicefrac}
\usepackage[titletoc,title]{appendix}

\renewcommand{\vec}[1]{\boldsymbol{ #1 }}
\newcommand{\ket}[1]{\left| #1 \right>}
\newcommand{\bra}[1]{\left< #1 \right|}
\newcommand{\amp}[2]{\langle  #1 | #2 \rangle}
\newcommand{\abs}[1]{\left| #1 \right|}
\newcommand{\e}{\mathrm{e}}
\newcommand{\ketbra}[2]{\left| #1 \vphantom{#2} \right>\!\!\left< #2 \vphantom{#1} \right|}

\begin{document}
\title{Charge and spin textures of Ising quantum Hall ferromagnet domain walls}
\author{Jeroen Danon$^{1}$, Ajit C. Balram$^{2,3,4}$, Samuel S{\'a}nchez$^{2}$ and Mark S. Rudner$^{2,3}$}
\affiliation{$^1$Center for Quantum Spintronics, Department of Physics, Norwegian University of Science and Technology, NO-7491 Trondheim, Norway}
\affiliation{$^2$Center for Quantum Devices, Niels Bohr Institute, University of Copenhagen, 2100 Copenhagen, Denmark}
\affiliation{$^3$Niels Bohr International Academy, Niels Bohr Institute, University of Copenhagen, 2100 Copenhagen, Denmark}
\affiliation{$^4$The Institute of Mathematical Sciences, HBNI, CIT Campus, Chennai 600113, India}
\date{\today}

\begin{abstract}
We investigate the charge and spin structures associated with arbitrary smooth polarization textures in Ising (integer) quantum Hall ferromagnets.
We consider the case where the two polarizations (denoted `pseudospin' up and down) correspond to states with opposite physical spin and different Landau level indices, $n\uparrow$ and $m\downarrow$.
We derive analytic expressions for the charge and spin densities, as functions of the underlying pseudospin texture, and use these results to investigate different types of linear domain walls, both analytically and numerically.
We find that any smooth domain wall 
between two oppositely polarized domains carries a universal quantized charge dipole density proportional to the difference of Landau level indices, $n-m$.
Additionally, non-uniformities in the domain wall may give rise to excess net charge localized at the domain wall.
Interestingly, the  physical spin density associated with the domain wall generally exhibits a much more complex multipolar structure than that of the pseudospin texture.
These results should for example help to elucidate the mechanisms underlying nuclear electric resonance and nuclear polarization oscillations in Ising quantum Hall systems.
\end{abstract}
\maketitle

The quantum Hall effect provides a rich setting for exploring the physics of strongly correlated quantum many-body systems.
In addition to the topological transitions between distinct quantum Hall phases that occur between Hall conductance plateaus, interesting new symmetry breaking transitions may occur within a given plateau when additional degeneracies are present, e.g., due to spin, valley, layer, or orbital ``pseudospin'' degrees of freedom~\cite{Girvin00}.
The appearance of a quantized Hall plateau at filling factor $\nu = 1$ provides the simplest example of such ``quantum Hall ferromagnetism,'' as electron-electron interactions play a crucial role in opening a gap in the half-filled (spin-degenerate) lowest Landau level (LL)~\cite{Sondhi93}. 
While the exchange interaction responsible for $\nu = 1$ quantum Hall ferromagnetism respects full $SU(2)$ spin rotation symmetry, systems with pseudospin degeneracies may exhibit phase transitions with either reduced (e.g., Ising/easy-axis or XY/easy-plane~\cite{Wen92c, Yang94, Smet01, Muraki01, Jaroszynski02}) or enhanced (e.g., $SU(n)$ with $n > 2$~\cite{Lok04, Nomura06, Feldman09, Weitz2010, Cote10, Dean11, Young12, Feldman16, Sodemann17}) symmetries, depending on the physical nature and multiplicity of the degeneracies. 

In this work we study Ising-like (easy-axis) quantum Hall ferromagnets with a two-fold pseudospin degree of freedom.
Near the phase transition (as a function of the parameters that control the nominal pseudospin degeneracy), such systems may exhibit complex domain patterns and dynamics~\cite{Piazza99, DePoortere00, Jungwirth01, PhysRevB.66.041308, Smet01, Kumada08, Yusa04, Liu10, Lu17, Korkusinski17}.
Prominently, domains between spin-polarized and unpolarized variants of the fractional quantum Hall state at $\nu = 2/3$ have been shown to give rise to a variety of intriguing dynamical phenomena including self-oscillations~\cite{Yusa04,Yusa05,Hennel16} and nuclear electric resonance~\cite{Kumada08, Watanabe10, Watanabe12, Korkusinski17}, and have even been proposed as a platform for realizing parafermions in hybrid structures involving superconductors~\cite{Wu18}.

\begin{figure}[t!]
\centering
\includegraphics[width=1.0\columnwidth]{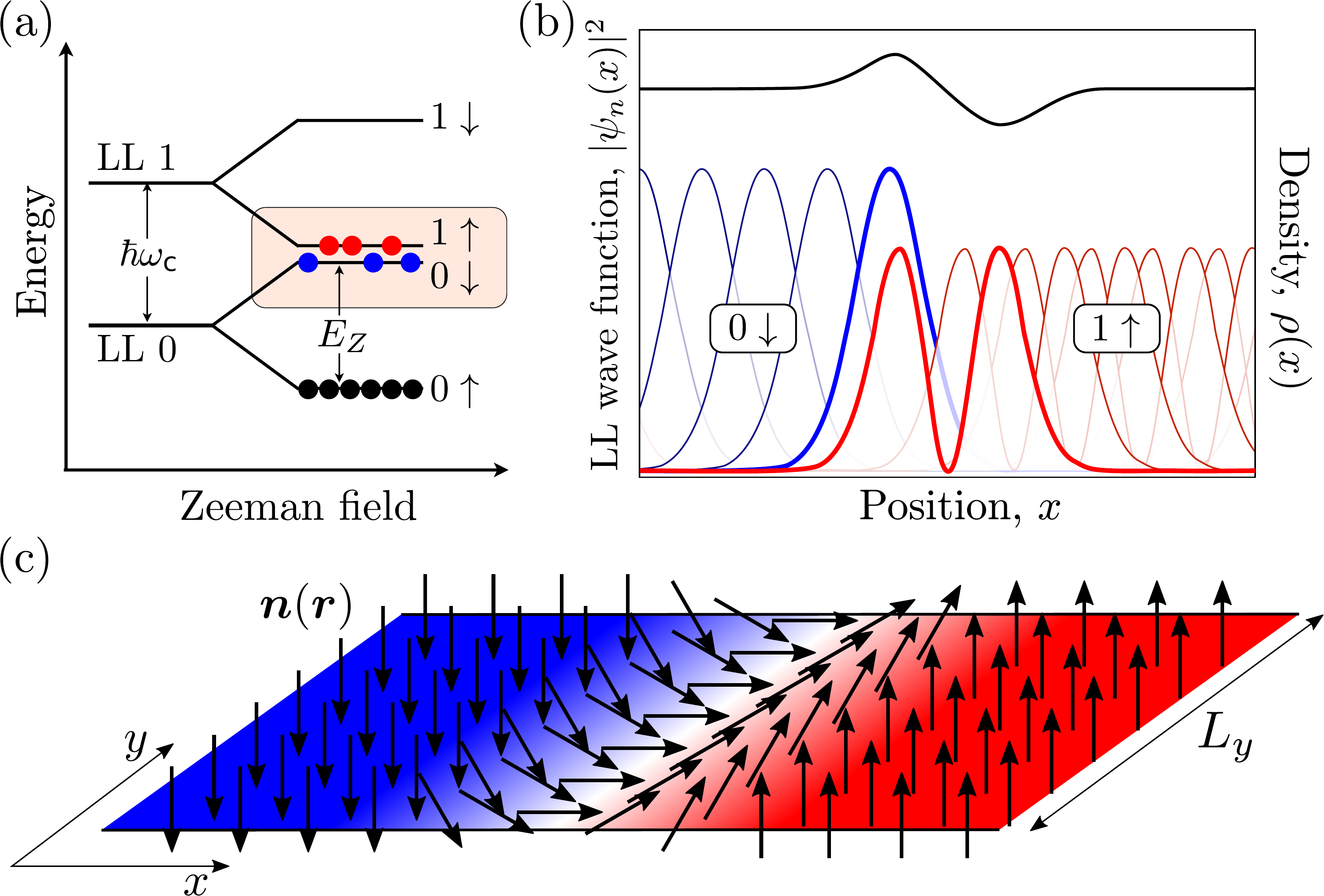}
\caption{Anatomy of a domain wall in an Ising quantum Hall ferromagnet. 
(a) Schematic energy spectrum. 
When the Zeeman energy $E_Z$ is approximately equal to the cyclotron energy $\hbar \omega_c$, the first LL with spin-up, denoted $1\uparrow$ and the zeroth LL with spin-down, denoted $0\downarrow$, are nearly degenerate. 
When this (highlighted) two-LL subspace is half-filled, interactions favor full polarization in the spin-LL basis.
(b) A dipole forms in the charge density (black line) at a domain wall between $0\downarrow$ and $1\uparrow$  regions due to the overlaps of the different Landau level wave functions (blue and red) from the two regions where they meet at the domain wall.
(c) Schematic of a linear domain wall along the $y$-direction, separating regions fully polarized in $0\downarrow$ (blue, arrows pointing down) and in $1\uparrow$ (red, arrows pointing up).}
\label{fig:summary}
\end{figure}

Below we characterize the charge and (physical) spin textures that naturally accompany domain walls in Ising-type quantum Hall ferromagnets in the integer quantum Hall effect regime.
We focus on the subspace comprised of two energy-degenerate Landau levels (LLs) of opposite spin, where the LL corresponding to spin-up ($\uparrow$) has the LL index $n$ and the one corresponding to spin-down ($\downarrow$) has the LL index $m$ (see Fig.~\ref{fig:summary}a, where the case $n = 1$, $m = 0$ is illustrated)~\cite{footnote:spin_only}.
Such a nearly-degenerate level configuration can be achieved by tuning the cyclotron and spin-splitting scales independently, using the fact that the cyclotron energy $\hbar\omega_c$ depends only on the perpendicular component of the external magnetic field, $B_{\perp}$, while the Zeeman energy $E_Z$ depends on the \textit{total} magnetic field, $B_{\rm tot}$~\cite{Jungwirth98}.
The resulting two-fold mixed orbital-spin degree of freedom ($n\uparrow$ and $m\downarrow$) that distinguishes states in the degenerate subspace forms the pseudospin for our system. 
Note that the field value where nominal degeneracy is achieved between the pseudospin-up states ($n\uparrow$) and pseudospin-down states ($m\downarrow$) is determined not only by the competition between the cyclotron and Zeeman energies, but also includes a contribution from intra-LL interaction energy (which may differ for LLs $n$ and $m$).

When the degenerate subspace outlined above is half-filled, electron-electron interactions lead to a ferromagnetic pseudospin ordering: either the pseudospin-up (LL $n$ with spin up) or pseudospin-down (LL $m$ with spin down) manifold is fully-filled, while ordering does {not} occur in any coherent superposition of the two. 
This easy-axis Ising ferromagnetic character results from two factors~\cite{Jungwirth2001a}:
First, due to the exchange energy, states with maximum total physical spin (as opposed to pseudospin) are naturally energetically preferred.
Second, due to the different orbital wave functions for states in LLs $n$ and $m$, inter-orbital interactions favor full orbital polarization.
Near the transition point between the fully polarized pseudospin-up and pseudospin-down phases, domain walls between spatial regions with opposite polarization can arise due to inhomogeneities of the sample, of the magnetic or Overhauser (hyperfine) fields, or any other local effect which favors one spin-polarization over the other~\cite{Jungwirth01,Fertig94,Fertig97,MacDonald97, Falko99, Mitra03, Kazakov17,Wu18}.

We analyze the case of a simple linear domain wall, and find via direct analytical and numerical calculation that the LL orbital structure associated with the pseudospin (see Fig.~\ref{fig:summary}b) generically induces a local electric dipole moment transverse to the domain wall.
The dipole moment per unit length along the domain wall, i.e., the dipole density, includes a universal, quantized, part, that depends only on the difference of LL indices, $n - m$, and arises for any smooth pseudospin texture that interpolates between pure pseudospin-up and pseudospin-down regions. 
The existence of this quantized contribution to the dipole density can also be traced to the difference in Hall viscosities associated with the two phases~\cite{Haldane09,Park14}.
We additionally find a nonuniversal contribution to the electric charge density that results from ``twisting'' of the pseudospin along the domain wall~\cite{Jungwirth01}.

In addition to the local charge density, we also characterize the physical spin texture within the domain wall.
Characterizing the spin texture is crucial for elucidating the coupling between domain wall degrees of freedom and nuclear spins in the host lattice~\cite{Hashimoto02, Yusa04, Hennel16, YangK17}, in particular for example in nuclear electric resonance experiments~\cite{Kumada08, Miyamoto16, Korkusinski17}.
Due to the different orbital wave functions within LLs $n$ and $m$, we find that the in-plane components of the total electron spin polarization generally vanish.
Locally, a nonvanishing in-plane spin {\it density} may appear within the domain wall, exhibiting a multipolar structure across the domain wall.

The structure of the paper is as follows.
In Sec.~\ref{sec:analytic} we explicitly define pseudospin density operators for systems with arbitrary LL indices, and use these operators to analytically derive expressions for the local deviations of charge and spin densities in the long wavelength limit of a smooth domain wall profile.
Then in Sec.~\ref{sec:Numerics} we numerically investigate the charge and spin textures for generic domain wall profiles for the case $n = 1$, $m = 0$, as depicted in Fig.~\ref{fig:summary}.
Finally, in Sec.~\ref{sec:Discussion} we provide a further discussion of our results.

\section{Definitions and analytical results}
\label{sec:analytic}

In this section we formalize the definitions of pseudospin and pseudospin density, and we derive analytical relations between pseudospin texture and local charge and spin densities that are valid in the limit of smooth polarization textures.

\subsection{Definition of pseudospin density}

We consider a system in which all spin-up LLs with index less than $n$ and spin-down LLs with index less than $m$ are fully occupied, while all spin-up (spin-down) LLs with indices greater than $n$ ($m$) are completely empty.
Without loss of generality we take $n>m$. 
Under the assumption that fully occupied and completely empty LLs are inert, we focus solely on the manifold comprised of the LLs $n\uparrow$ and $m\downarrow$ (highlighted in Fig.~\ref{fig:summary}a for the case $n = 1, m = 0$), which we assume to be half filled. 

For concreteness we study the system on a strip lying in the $xy$-plane, where $-\infty<x<\infty$ and $0<y<L_y$.
Appropriate to this geometry, we will use the Landau gauge, assuming periodic boundary conditions in the $y$-direction for convenience.
We thus use the explicit single-particle basis states 
\begin{align}
\langle{{\bf r}|n,k}\rangle  = {} & {} \frac{e^{-iky}}{\sqrt{L_y 2^n n! \ell \sqrt \pi }}
\, H_n \! \left( \frac{x - k\ell^2}{\ell} \right) e^{ - \frac{ (x-k\ell^2)^2 }{ 2\ell^2 }}, \label{eq:llwavefunc}
\end{align}
labeled by the LL index $n$ and the quantum number $k$ corresponding to the canonical momentum in the $y$-direction.
Here $\ell = \sqrt{\hbar/eB_\perp}$ is the magnetic length and $H_n(x)$ denotes the $n$-th Hermite polynomial.

In first quantization, for a single particle, $j$, we define the pseudospin $\boldsymbol{\tau}_j$ within the $\{n\uparrow, m\downarrow\}$ LL subspace via
\begin{align}
\label{eq:tau0}  \tau^{0}_j & = 
            \sum_k 
            \begin{pmatrix}
            \ketbra{n,k}{n,k} & 0\\
            0 & \ketbra{m,k}{m,k}
            \end{pmatrix}\,,
            \\
\label{eq:taux} \tau^{x}_j & =
            \sum_k
            \begin{pmatrix}
            0 & \ketbra{n,k}{m,k}\\
            \ketbra{m,k}{n,k} & 0
            \end{pmatrix}\,,
            \\
\label{eq:tauy} \tau^{y}_j & =
            \sum_k
            \begin{pmatrix}
            0 & -i\ketbra{n,k}{m,k}\\
            i\ketbra{m,k}{n,k} & 0
            \end{pmatrix}\,,
            \\
\label{eq:tauz} \tau^{z}_j & = 
            \sum_k
            \begin{pmatrix}
            \ketbra{n,k}{n,k} & 0\\
            0 & -\ketbra{m,k}{m,k}
            \end{pmatrix}\,,
\end{align}
where $\ket{\alpha,k}$ (with the particle index $j$ suppressed) denotes a state in LL $\alpha = \{n, m\}$, and
the $2\times 2$ matrices act in physical spin space $\{\uparrow,\downarrow\}$.
The $\vec{\tau}_j$ operators fulfil the usual spin Lie algebra, and can serve as generators of global rotations in pseudospin space. 

To enable the mathematical creation and characterization of \textit{inhomogeneous} pseudospin textures, we upgrade the global (single-particle) pseudospin operators in Eqs.~(\ref{eq:tau0})--(\ref{eq:tauz}) to {\it local} (many-body) pseudospin density operators:
\begin{align}
\label{eq:pseudospindensity-unproj}
\tau^{\eta}(\vec{r}) = \sum_j \frac{1}{2} \left[ \delta(\hat{\vec{r}}_j - \vec{r})\tau_j^{\eta} +  \tau^{\eta}_j \delta(\hat{\vec{r}}_j-\vec{r}) \right], 
\end{align}
where $\eta=0,x,y,z$ and $\hat{\vec{r}}_j$ is the position operator for electron $j$ and the sum is over all particles.
The symmetrization of the product with the delta function ensures that the operator is Hermitian. 
We note that the operator $\delta(\hat{\vec{r}}_j-\vec{r})$ in Eq.~(\ref{eq:pseudospindensity-unproj}) can take electron $j$ out of the $\{{n\uparrow}, {m\downarrow}\}$ subspace.
Since we are interested in creating states restricted to this low energy subspace, we thus work instead with the {\it projected} pseudospin density
\begin{equation}\label{eq:pseudospindensity-proj}
\overline{\boldsymbol{\tau}}(\vec{r}) = \mathcal{P}\, \boldsymbol{\tau}(\vec{r})\, \mathcal{P}, \quad \mathcal{P} = \bigotimes_j \tau^{0}_j,
\end{equation}
where we used that $\tau^{0}_j$ corresponds to the projector into the $\{{n\uparrow}, {m\downarrow}\}$ manifold of interest for electron $j$, see definition in Eq.~(\ref{eq:tau0}).

We emphasize an important difference between pseudospin and physical spin, which we illustrate by considering a general single-particle state in the $\{{n\uparrow}, {m\downarrow}\}$ subspace, $\ket{\psi}~=~\sum_k \left(a_k^n\ket{n,k}\otimes\ket{\uparrow} + a_k^m\ket{m,k}\otimes\ket{\downarrow}\right)$, where $\sum_{k}(|a_k^n|^{2}+|a_k^m|^{2})=1$.
Due to the orthogonality of the LL states with $n\neq m$, only the $z$-component of the physical spin operator may have a non-vanishing \emph{total} expectation value. 
From the definitions above, it is straightforward to see that the $z$-components of the pseudospin and of the physical spin in fact have identical expectation values. 
The situation with the in-plane ($x$ and $y$) components is more subtle.
The in-plane components of the physical spin can only have nonvanishing \emph{local} expectation values of the corresponding spin densities. 
On the other hand, all the components of the pseudospin can have nonvanishing global and local expectation values. 

\subsection{Pseudospin textures and charge density}\label{sec:charge}

Here we investigate how the charge density in the system is influenced by inhomogeneities in the pseudospin density (as in a domain wall). 
To calculate the change in charge density, we follow a generalized version of the procedure of Ref.~\onlinecite{Moon95}, where the special case of $n=m=0$ was considered. 
In that work, the authors considered deformations around a uniformly polarized state in which all electron spins point in the positive $z$-direction. 
They constructed a unitary rotation operator that produces deformations on top of this state, and found the resulting excess charge density for small deformations by expanding the deformation operator up to second order in the deformation amplitude. 
Using $SU(2)$ symmetry they inferred the full spin-charge relation (to leading order in gradients of the polarization vector) and showed that the global excess charge carried by the spin deformation is quantized in units of the filling factor. 

A major difference between the case considered by Moon \emph{et al.}~in Ref.~\onlinecite{Moon95} and ours is that, in our problem, $SU(2)$ symmetry is broken down to $U(1)$ due to the Ising-like nature of the interactions.
Therefore, (i) we do not expect the pseudospin-charge relation to exhibit full $SU(2)$ symmetry, and (ii) we cannot simply consider small deformations around a single initial state (say, with all spins pointing to $+z$), and then infer how the local charge density is affected for deformations around all other directions of pseudospin polarization.

What we will do instead is consider an initial homogeneous state $|\Psi_0\rangle$ where all pseudospins are polarized along $\vec{n}_0 = (\sin \theta\cos\phi, \sin\theta\sin\phi, \cos\theta)$, where $\theta$ and $\phi$ are the polar and azimuthal angles of the polarization direction, respectively.
We take the many-body state $\ket{\Psi_0}$ to be a Slater-determinant wave function over all particles, exactly filling one particle per $k$ mode, with the single-particle state for each $k$ being $\ket{\psi_{k}} = \cos(\theta/2)\ket{n,k}\otimes\ket{\uparrow} + e^{i\phi} \sin(\theta/2)\ket{m,k}\otimes\ket{\downarrow}$.

We proceed by applying a small position-dependent pseudospin rotation to deform the initial homogeneous state, using the pseudospin rotation operator
\begin{align}
\label{eq:R}
U_R=\e^{-i \overline{O}},\quad\text{with}\quad
\overline{O}=
\int\!\!d\vec{r}\ \frac{1}{2}\,\vec{\Omega}(\vec{r}) \cdot \overline{\vec{\tau}}(\vec{r}),
\end{align}
where  $\vec{\Omega}(\vec{r})$ defines the axis and angle over which the pseudospin is rotated from its initial orientation.
We note here that only in the long wavelength limit of smooth $\vec{\Omega}(\vec{r})$ (on the scale of $\ell$) can the local pseudospin density at point $\vec{r}$ be obtained precisely by rotating the initial pseudospin through an angle $\Omega(\vec{r})$ about the axis parallel to $\vec{\Omega}(\vec{r})$~\cite{footnote:rotations}.
(Here and throughout we use boldface to denote vectors and normal fonts their magnitudes.)
We will employ this long wavelength limit in our analytic calculations.

We now turn to calculating the change in the expectation value of the projected charge density $\overline{\rho}(\vec{r}) \equiv \overline{\tau}^0(\vec{r})$ induced after applying the rotation $U_R$  (for convenience we will use $\hbar = e = 1$ throughout):
\begin{equation}
\langle \delta\overline{\rho}_{\vec{q}} \rangle =
            \bra{\Psi_0}{
                \e^{i\overline{{O}}}
        \overline{\rho}_{\vec{q}}
        \e^{-i\overline{{O}}}}\ket{\Psi_0}
        -
        \bra{\Psi_0}{
            \overline{\rho}_{\vec{q}}}\ket{\Psi_0},
\label{eq:exact-change-charge-density}            
\end{equation}
where $\overline{\rho}_{\vec{q}}$ is the Fourier transform of $\overline{\rho}(\vec{r})$, see App.~\ref{appendix_explicit_definitions}. 
Assuming the limit of small rotation angles, $||\overline{O}||\ll 1$, we expand $U_R$ to second order in $\overline{O}$ to obtain            
\begin{align}
\label{eq:expanded_rho}
\langle \delta\overline{\rho}_{\vec{q}} \rangle \approx {} & {}
        i\bra{\Psi_0}{
            [\overline{{O}},\overline{\rho}_{\vec{q}}
            ]
            }
            \ket{\Psi_0}
            -\frac{1}{2}\bra{\Psi_0}{
            [\overline{{O}},
            [\overline{{O}},\overline{\rho}_{\vec{q}}
            ]
            ]
            }
            \ket{\Psi_0}
            .
\end{align}

To begin, we consider the first-order term in (\ref{eq:expanded_rho}).
Using the Fourier representation of $\overline{{O}}$, as defined in Eq.~(\ref{eq:R}), we find for the first-order correction
    \begin{align}
        \langle \delta\overline{\rho}_{\vec{q}}^{(1)} \rangle &=
i\pi L_y 
\sum_{\eta=x,y,z}\sum_{\vec{p}}
{\Omega}^{\eta}_{\vec{p}} 
\bra{\Psi_0}{
	[
	\overline{\tau}^{\eta}_{-\vec{p}},
	\overline{\rho}_{\vec{q}}
	]
}\ket{\Psi_0}.
    \end{align}
Here, ${\Omega}^{\eta}_{\vec{p}}$ and $\overline{\tau}^{\eta}_{-\vec{p}}$ are the Fourier transforms of $\Omega^{\eta}(\vec{r})$ and $\overline{\tau}^{\eta}(\vec{r})$, respectively.
Using the explicit forms of the commutators $[\overline{\tau}^{\eta}_{-\vec{p}},\overline{\rho}_{\vec{q}}]$ in the long wavelength limit $q\ell \ll 1$ we find
    \begin{align}
         \langle \delta\overline{\rho}_{\vec{q}}^{(1)} \rangle 
\label{eq:rho_q_final} & =
         \frac{n-m}{8\pi} \abs{\vec{q}}^2 \sin\theta
         \big[ \cos\phi\, \Omega^{y}_{\vec{q}} - \sin\phi\, \Omega^{x}_{\vec{q}}\big],
    \end{align}
up to corrections that are smaller by a factor of the order $\mathcal{O}[(q\ell)^2]$.
We note that this (first-order) change in charge density vanishes for $n = m$, which is in accordance with the results of Ref.~\onlinecite{Moon95}.
Fourier transforming the charge density back to real space, we find
\begin{align}
        \langle \delta\overline{\rho}^{(1)}(\vec{r}) \rangle 
       =
        \frac{n-m}{8\pi} \nabla^2_{\vec{r}} \sin\theta
        \big[\sin\phi\, \Omega^{x}(\vec{r}) - \cos\phi\, \Omega^{y}(\vec{r}) \big].\label{eq:dr1}
\end{align}

We elucidate the structure of this expression by considering the normalized vector field $\vec{n}(\vec{r}) = R[\boldsymbol \Omega(\vec{r})] \cdot \vec{n}_0$, where $R[{\bf u}]$ is the three-dimensional rotation operator that implements a rotation through the angle $u$ around the axis $\vec{u}/u$.
By definition (in the long wavelength limit), $\vec{n}(\vec{r})$ then points along the local pseudospin polarization everywhere: $\vec{n}(\vec{r}) \parallel \langle \overline{\boldsymbol{\tau}}(\vec{r})\rangle$.
By considering the explicit form of $R[\boldsymbol \Omega(\vec{r})]$, see Eq.~(\ref{eq:3drot}), we find that Eq.~(\ref{eq:dr1}) coincides with the expression
\begin{equation}
\langle \delta\overline{\rho}^{(1)}(\vec{r})\rangle =\frac{n-m}{8\pi}\,\nabla^2_{\vec{r}}n^z(\vec{r}),
\label{eq_particle_density}
\end{equation}
up to first order in $\Omega$.

The procedure above can be continued for the second-order term in Eq.~(\ref{eq:expanded_rho}).
The algebra (not shown here) is long and unilluminating, but in the end reveals two separate contributions:
(i) we precisely obtain a contribution that adds to  Eq.~(\ref{eq:dr1}) in a way that the resulting total correction is equivalent to  Eq.~(\ref{eq_particle_density}) up to second order in $\Omega$,
and (ii) we recover a fully $SU(2)$-symmetric ``Pontryagin index density,'' identical to the one found in Ref.~\onlinecite{Moon95} for the case $n=m$.
Thus, written explicitly, we obtain up to second order in pseudospin gradients
\begin{equation}
\label{eq:spin-charge-final} 
\langle \delta\overline{\rho}(\vec{r}) \rangle = \frac{n-m}{8\pi}\nabla^2_{\vec{r}}n^z + \frac{\epsilon_{\alpha\beta}}{8\pi} \vec{n} \cdot [\partial_\alpha \vec{n} \times \partial_\beta \vec{n}],
\end{equation}
where the antisymmetric tensor $\epsilon_{\alpha\beta}$ has components $\epsilon_{xy} = -\epsilon_{yx} = 1$ and is zero otherwise, and we suppress the position arguments of $\vec{n}$ for brevity.
Note that the second term in this expression does not depend on the LL indices of the states involved.

From a symmetry point of view, it is perhaps surprising that we recover the fully $SU(2)$-symmetric from of the Pontryagin index density, given that the system has only $U(1)$ symmetry.
In particular, based on symmetry alone, a contribution of the form $\delta \overline{\rho}^{(2)} \sim \epsilon_{\alpha\beta}\, n^z\, \hat{\vec{z}}\cdot[\partial_\alpha \vec{n} \times \partial_\beta \vec{n}]$ could in principle appear with a different prefactor from the other terms.

Equation (\ref{eq:spin-charge-final}) was derived by considering small rotations away from an initial polarization along the specific direction set by $\vec{n}_0$ [see text above Eq.~(\ref{eq:R})].
In the limit of smooth textures considered here, where $q\ell \ll 1$ is always satisfied, we can expect Eq.~(\ref{eq:spin-charge-final}) to hold globally as well (essentially by locally resetting $\vec{n}_0$ at each $\vec{r}$).

\subsection{Quantized contribution to the dipole density at a domain wall}\label{sec:dipole}

Now we consider a simple linear domain wall parallel to the $y$-direction, separating domains of opposite pseudospin polarization, i.e., we assume that $\vec{n} = - \hat{\vec{z}}$ at $x \to -\infty$ and $\vec{n} = + \hat{\vec{z}}$ at $x \to +\infty$, see the sketch in Fig.~\ref{fig:summary}c.
We furthermore specify the domain wall to be created by a uniaxial rotation, $\vec{\Omega}({\vec{r}}) = f(x)\,\hat{\vec{\zeta}}$, where $\hat{\vec{\zeta}}$ can be any in-plane direction.
In this case, the second term in Eq.~(\ref{eq:spin-charge-final}) vanishes everywhere. 

To find the total charge associated with the domain wall, we then integrate Eq.~(\ref{eq:spin-charge-final}) over all $\vec{r}$.
Assuming only that the pseudospin polarization is uniform in the asymptotic regions, i.e., $\nabla_{\vec{r}}n^z(\vec{r}) = 0$ for $x \rightarrow \pm \infty$, we find straightforwardly that the total charge associated with the domain wall vanishes.
Note that this result is insensitive to the details of the profile of $f(x)$.

The dipole moment density $\mu$, defined as the dipole moment in the $x$-direction per unit length along the $y$-direction, is obtained by integrating $\langle \delta\overline{\rho}(\vec{r}) \rangle$ in Eq.~(\ref{eq:spin-charge-final}) along $x$, weighted by the value of the $x$-coordinate.
For a uniaxial rotation, and in the long wavelength limit, integration by parts gives directly
\begin{equation}
  \mu = \int_{-\infty}^{\infty}\!\!dx\,x\langle \delta\overline{\rho}(\vec{r}) \rangle=
-\frac{1}{4\pi}(n-m).
\label{eq_linear dipole_moment_density}
\end{equation}
Hence we find that the dipole moment per unit length in a linear domain wall created by uniaxial rotations is quantized and universal: it depends only on the difference of the indices of the LLs involved, and does not depend on the precise form of the domain wall set by $f(x)$.

On a qualitative level, the appearance of an electric dipole at a domain wall between integer quantum Hall states with different LL indices can be intuitively understood by considering the overlaps between neighboring LL wave functions~(see Fig.~\ref{fig:summary}b). 
At the domain wall, the wave functions corresponding to spin-up (LL $n$) have a larger spread than the wave functions corresponding to spin-down (LL $m$). 
As a consequence, charge density ``leaks'' from the spin-up region into the spin-down region, resulting in a build-up of excess charge density on one side of the domain wall and a deficit on the other~\cite{footnote:uniform_density}.

We note that the result of Eq.~(\ref{eq_linear dipole_moment_density}) is consistent with the previously-known result that the electric dipole moment per unit length induced at the boundary of two quantum Hall fluids is directly proportional to the difference in their Hall viscosities~\cite{Haldane09,Park14, footnote:HallViscosity}. 

\subsection{Pseudospin textures and spin density}

Following a similar procedure as outlined in Sec.~\ref{sec:charge}, we also calculate the physical spin density associated with a smooth pseudospin texture.
The (many-body) spin density operators read
\begin{equation}
s^{\eta}(\vec{r}) = \sum_j\delta(\hat{\vec{r}}_j - \vec{r}) \frac{1}{2}\sigma^{\eta}_j,
\end{equation}
where the $\{\sigma^{\eta}_j\}$ are the usual Pauli matrices for $\eta = x,y,z$.
No symmetrization is needed here since the spin operators and the particle density operator commute [cf.~Eq.~(\ref{eq:pseudospindensity-unproj})].
Projecting to the subspace $\{ n\uparrow, m\downarrow \}$ using the operator $\mathcal{P}$ defined in Eq.~(\ref{eq:pseudospindensity-proj}), we obtain the projected spin density:
\begin{equation}
\overline{\boldsymbol s}(\vec{r}) = \mathcal{P}\, {\boldsymbol s}(\vec{r})\, \mathcal{P}.
\end{equation}

Starting from the same initial state $\ket{\Psi_0}$ with a homogeneous pseudospin polarization as used above, see text before Eq.~(\ref{eq:R}), we again apply a small position-dependent rotation as defined in (\ref{eq:R}).
Up to first order in $\overline{O}$, the Fourier transformed change in spin density follows from
\begin{equation}
\langle \delta \overline{s}^{\eta}_{\vec{q}} \rangle \approx 
i\bra{\Psi_0}{
	[\overline{{O}},\overline{s}^{\eta}_{\vec{q}}
	]
}
\ket{\Psi_0}.
\end{equation}

A tedious but straightforward calculation then allows us to derive explicit results.
Since the  $z$-components of the projected spin and pseudospin density operators are identical (up to a factor $\frac{1}{2}$), $\langle \overline{s}^{z}_{\vec{r}} \rangle$ simply follows the pseudospin polarization texture.
For the simplest case of $n=1$ and $m=0$, we find for the in-plane spin densities: 
\begin{align}
\langle \delta \overline{s}^x_{\vec{q}} \rangle
= \frac{i\sqrt 2}{8\pi \ell}  \big\{  {} & {}  \cos \theta [q_x \Omega^y_{\vec{q}} -q_y \Omega^x_{\vec{q}} ] 
\nonumber\\ {} & {}
+\sin\theta  [\cos\phi\,q_y-\sin\phi\, q_x ] \Omega^z_{\vec{q}} \big\},\label{eq:sigmax_q} \\
\langle \delta \overline{s}^y_{\vec{q}} \rangle 
= \frac{i\sqrt 2}{8\pi \ell}  \big\{  {} & {}  - \cos \theta [q_x \Omega^x_{\vec{q}} +q_y \Omega^y_{\vec{q}} ] 
\nonumber\\ {} & {}
+\sin\theta  [\cos\phi\,q_x+\sin\phi\, q_y ] \Omega^z_{\vec{q}} \big\},\label{eq:sigmay_q}
\end{align}
again valid in the limit of long wave lengths, $q\ell \ll 1$ and accurate up to relative corrections of order ${\cal O}[(q\ell)^2]$.

We find expressions similar to Eqs.~(\ref{eq:sigmax_q}) and (\ref{eq:sigmay_q}) for the case of general $n$ and $m$; Fourier transforming them back to position space, while using that the gradients of $\boldsymbol\Omega(\vec{r})$ are assumed to be small, we can simplify the final result to
\begin{align}
\langle \delta \overline{s}^{+}(\vec{r}) \rangle
{} & {} = \frac{1}{4\pi \ell^2}
\frac{\sqrt{n!/m! } }{(n-m)!} 
(-\tilde \partial_-)^{n-m} n^+(\vec{r}),\label{eq:spin_full}
\end{align}
where we introduced the notation $\overline{s}^{+}(\vec{r}) = \overline{s}^{x}(\vec{r})+i\overline{s}^{y}(\vec{r})$,  $n^+(\vec{r}) = n^x(\vec{r}) + in^y(\vec{r})$, and $\tilde \partial_- =\frac{\ell}{\sqrt 2} ( \partial_x - i\partial_y)$.
Again, these expressions are accurate up to relative corrections of the order ${\cal O}[(q\ell)^2]$, i.e., they assume the texture $\vec{n}(\vec{r})$ to be smooth on the scale of $\ell$.

From Eq.~(\ref{eq:spin_full}) we see that domain walls in general also can give rise to non-trivial structure in the in-plane spin densities.
For the case of $n=1$ and $m=0$ it is easy to show that a simple domain wall, such as one created by a uniaxial rotation (cf.~Sec.~\ref{sec:dipole}), in general comes with an in-plane spin polarization with a dipole-like structure close to the domain wall.
Naively, one might have expected the spin density to simply follow the pseudospin texture (which typically does {\it not} have a dipolar structure), since the pseudospin up and down states also carry physical up and down spins, respectively.
However, the in-plane spin density is locally sensitive to the relative phases between single-particle wave functions $\psi_{0,k}(\vec{r}) = \amp{\vec{r}}{0, k}$ and $\psi_{1,k}({\vec{r}}) = \amp{\vec{r}}{1, k}$ associated with down and up spins, respectively.
As apparent in Eq.~(\ref{eq:llwavefunc}), each function $\psi_{1,k}(\vec{r})$ is odd in $x$ (with respect to a $k$-dependent symmetry axis), while $\psi_{0,k}(\vec{r})$ is even in $x$ about the same symmetry axis.
Therefore the overlap $\psi^*_{0,k}(x)\psi_{1,k}(x)$ is odd in $x$, and hence the in-plane spin component takes opposite signs on opposite sides of the symmetry axis.
The main contribution to the in-plane spin density, coming from the center of the domain wall where the pseudospin up and down components have equal weight, is thus antisymmetric.
This results in a net dipole-like structure for the in-plane component of the spin density.
We note that the appearance of a simple dipole in the spin density is not a generic feature, but specific for the case $n=1$ and $m=0$; other $n$ and $m$ result in different, possibly more complex textures, see Eq.~(\ref{eq:spin_full}).

\section{Numerical results}\label{sec:Numerics}

To further elucidate our analytical results, we now present numerical calculations of the charge and spin densities around different domain wall structures.
For simplicity we will focus on the case with $n=1$ and $m=0$, but we have checked that our results are also valid for other values of $n$ and $m$.

\subsection{Uniform domain wall}\label{sec:numuni}

We start by creating a simple homogeneous domain wall, resembling the form found in Ref.~\cite{Jungwirth01} from self-consistently solving the Hartree-Fock equations~\cite{footnote:no_minimization}.
To that end, we first create a fully pseudospin-up polarized $N$-particle Slater-determinant wave function, with one electron in each $k$ mode.
Specifically, we fill all states  $\ket{1,k}\otimes\ket{\uparrow}$ with $k = 2\pi l/L_{y}$, where $l$ is an integer in the range $-(N-1)/2\leq l \leq (N-1)/2$ (using odd $N$).
We then apply the rotation operator as defined in Eq.~(\ref{eq:R}) to all single-particle states, where we use the rotation function $\vec{\Omega}(\vec{r}) = f(x)\, \hat{\vec{y}}$, where $f(x)=\frac{\pi}{2}[\tanh(b\,x)-1]$, see App.~\ref{app:homodw} for details.
This choice describes a domain at $x\ll0$ where the system is pseudospin-polarized along the $-z$-direction and a domain at $x\gg0$ where the pseudospin is polarized in the $+z$ direction.
Around $x=0$ there will be a domain wall of width $\sim 1/b$ that is uniform in the $y$-direction.
The many-particle wave function resulting from this rotation operation can then be used to numerically calculate expectation values of all components of the (pseudo)spin and charge densities.

\begin{figure}[t!]
	\centering
	\includegraphics[scale=.48]{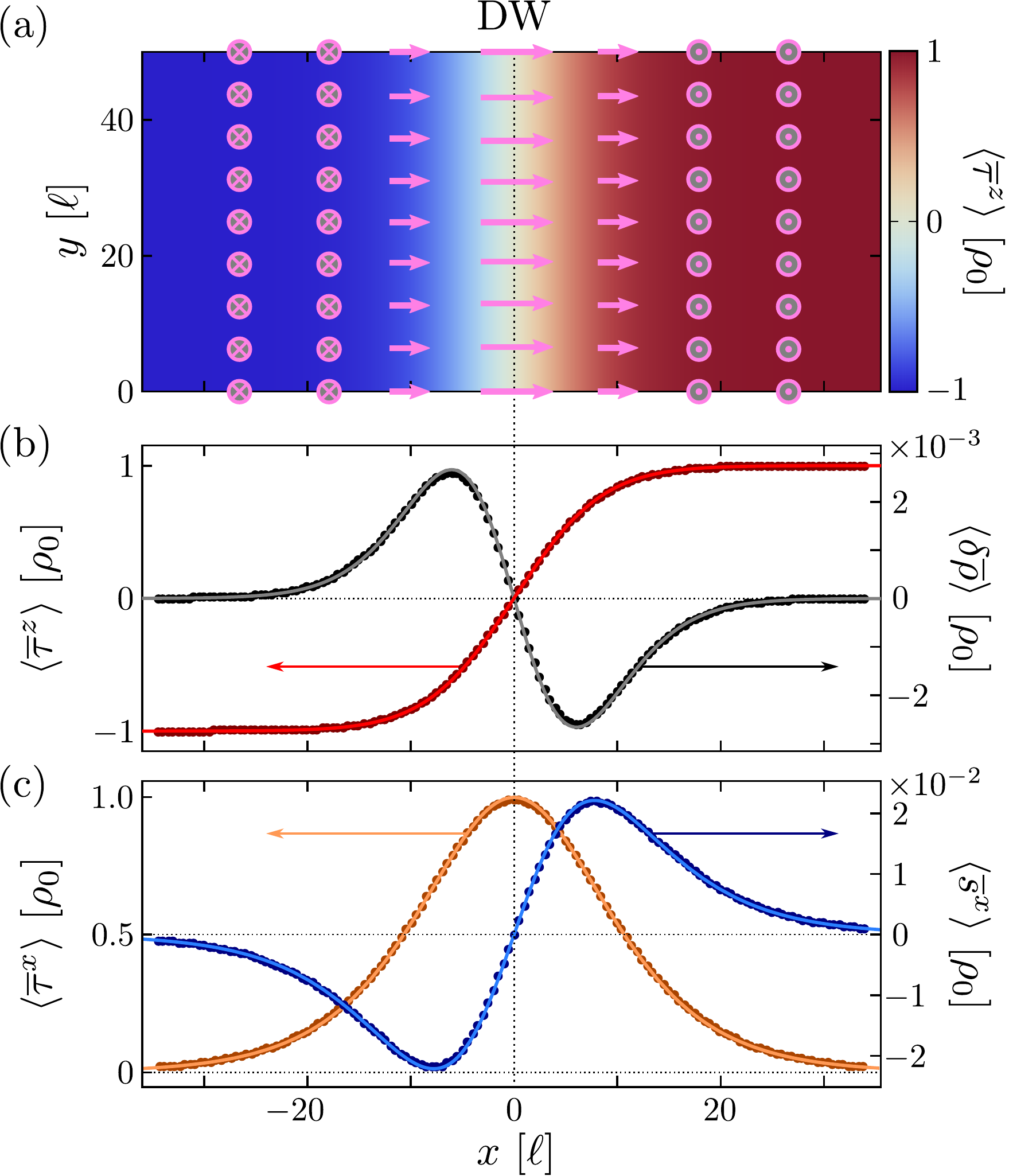}
	\caption{
		(a) Numerically calculated pseudospin density $\langle \overline{\tau}^z(\vec{r})\rangle$ after applying a rotation defined by $\vec{\Omega}(\vec{r})=\frac{\pi}{2}[\tanh(b\,x)-1]\, \hat{\vec{y}}$ with $b=0.075/\ell$.
		We further used $N=601$, $L_y = 50 \, \ell$.
		The pink arrows indicate the local direction of the vector $\langle \overline{\vec{\tau}}(\vec{r})\rangle$.
		(b,c) Dots: Line cuts at $y = 25\ell$ of numerically calculated values of (b) the $z$-component of the pseudospin density (red, left axis) and change in charge density (black, right axis) and (c) the $x$-component of the pseudospin (orange, left axis) and physical spin (blue, right axis) densities. Lines: Same densities calculated analytically, using $\langle \overline{\vec{\tau}}(\vec{r}) \rangle = \rho_0 \vec{n}(\vec{r})$, where $\vec{n}(\vec{r})= R[\vec{\Omega}(\vec{r})] \hat{\vec{z}}$ and $\rho_0$ is the background charge density, and $\langle \overline{s}^{x,y}(\vec{r}) \rangle$ and $\langle \delta \overline{\rho}(\vec{r})\rangle$ as given by Eqs.~(\ref{eq:spin_full}) and (\ref{eq:spin-charge-final}). All densities are plotted in units of $\rho_0$.
}
	\label{fig:num1}
\end{figure}

\begin{figure*}[ht!]
	\centering
	\includegraphics[scale=.48]{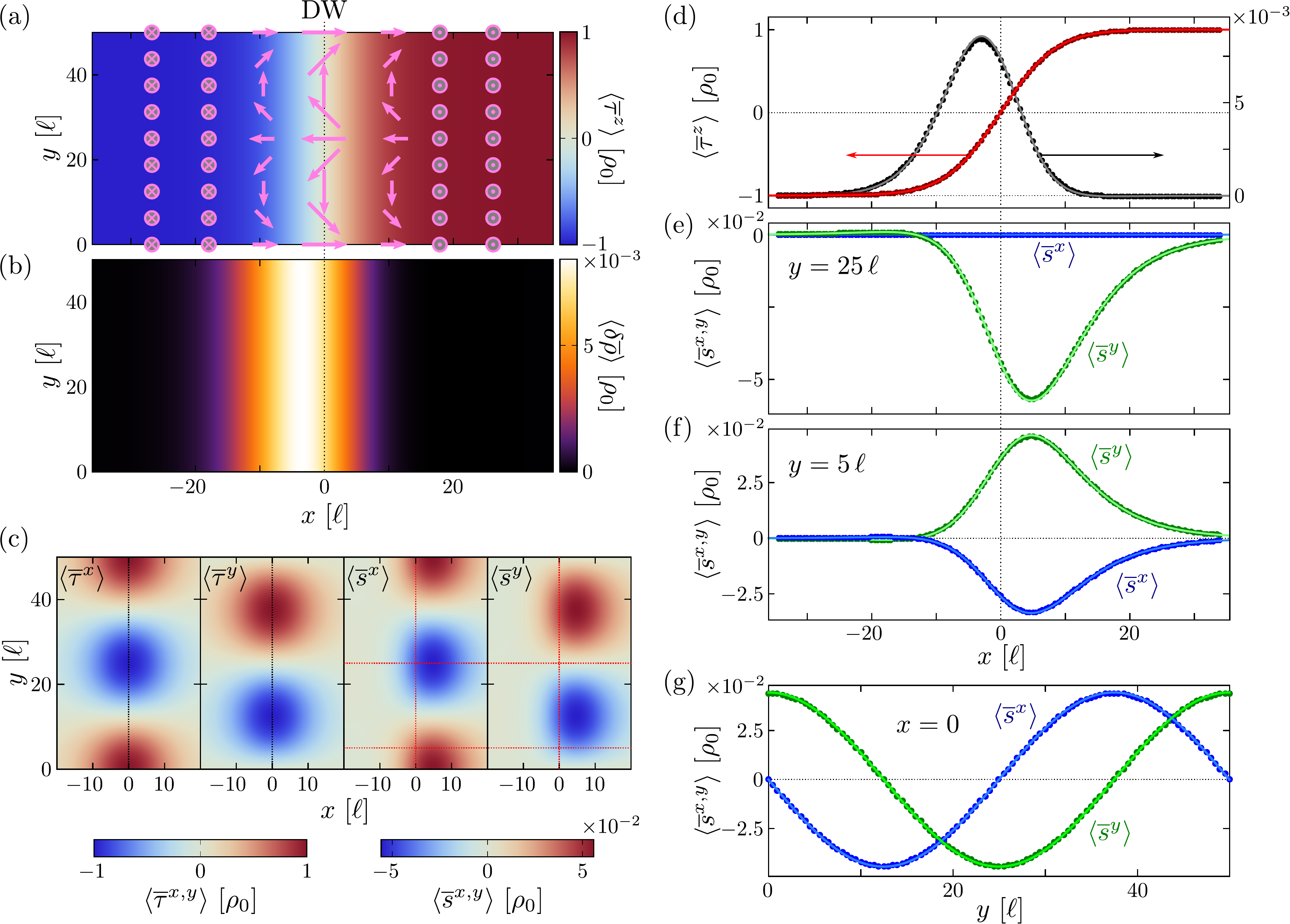}
	\caption{
		(a) Numerically calculated pseudospin density $\langle \overline{\tau}^z_{\vec{r}}\rangle$ after applying the same rotation as in Fig.~\ref{fig:num1}, but now with a $y$-dependent rotation axis $( \sin [2\pi y/L_y], \cos[2\pi y/L_y], 0 )$.
		The pink arrows show the in-plane pseudospin ``twist'' that is created.
		(b,c) Numerically calculated change in (b) charge density compared to the background density and (c) expectation values of the in-plane components of pseudospin and physical spin density.
		(d--g) Dots: Line cuts of numerically calculated values of expectation values shown in (a--c).
		Lines: Analytic results, using the same expressions as before.
		(d) $z$-component of pseudospin density (red, left axis) and change in charge density (black, right axis).
		(e--g) In-plane physical spin densities (green for $\langle \overline{s}^x\rangle$, blue for $\langle \overline{s}^y\rangle$), along (e) $y=25\,\ell$, (f) $y=5\, \ell$, and (g) $x=0$; these coordinates are indicated with the red dotted lines in (c).
All densities are again plotted in units of the background charge density, $\rho_0$.}
	\label{fig:num2}
\end{figure*}

In Fig.~\ref{fig:num1} we present results where we used $N=601$, $L_y = 50\,\ell$, and $b=0.075/\ell$.
Fig.~\ref{fig:num1}a shows the resulting domain wall structure; the color corresponds to the numerically calculated $z$-component of the pseudospin density $\langle \overline{\tau}^z(\vec{r}) \rangle$ and the pink arrows schematically indicate the direction of the vector $\langle \overline{\vec{\tau}}(\vec{r})\rangle$.
In Fig.~\ref{fig:num1}b, the black dots (right axis) present the numerically calculated change in charge density $\langle \delta\overline{\rho}(\vec{r}) \rangle = \langle \overline{\rho}(\vec{r})\rangle - \rho_0$ along $y= 25\, \ell$, where $\rho_0 = 1/(2\pi \ell^2)$ is the background charge density.
The gray line shows for comparison $\langle \delta \overline{\rho}(\vec{r})\rangle$ as given by Eq.~(\ref{eq:spin-charge-final}) using $\vec{n}(\vec{r}) = R[\vec{\Omega}(\vec{r})] \hat{\vec{z}}$, which agrees well with our numerical results.
We note that for this simple uniaxial rotation the contribution from the second term in Eq.~(\ref{eq:spin-charge-final}) is zero.
The red dots and line (left axis) show $\langle \overline{\tau}^z(\vec{r}) \rangle$ along $y = 25\,\ell$, both extracted from the numerical data (dots) and calculated using $\langle\overline{\vec{\tau}}(\vec{r})\rangle = \rho_0 \vec{n}(\vec{r}) = \rho_0 R[\vec{\Omega}(\vec{r})] \hat{\vec{z}}$ (line).
We see that the charge density indeed exhibits a dipole structure close to the domain wall, as predicted in Sec.~\ref{sec:dipole}. The dipole density calculated from these numerical results is $\mu = -0.07942$, which is very close to the predicted value of $-1/4\pi$.

In Fig.~\ref{fig:num1}c we plot the in-plane component of the pseudospin density $\langle \overline{\tau}^{x}(\vec{r}) \rangle$ (orange, left axis) and physical spin density $\langle \overline{s}^{x}(\vec{r}) \rangle$ (blue, right axis) along $y = 25\,\ell$.
The $y$-components of both densities are zero everywhere.
The dots present the numerically calculated values and the lines the analytical results, where $\langle \overline{\tau}^{x}(\vec{r}) \rangle$ was calculated as explained above and $\langle \overline{s}^{x}(\vec{r}) \rangle$ using Eq.~(\ref{eq:spin_full}).
Again we see good agreement between the analytic expressions and numerical results, and we confirm our prediction that the in-plane spin density associated with a simple domain wall also exhibits a dipole structure (qualitatively different from the in-plane {\it pseudospin} texture).

\subsection{Non-uniform domain wall}\label{sec:non-uni}

For further interest, we now create a more complex domain wall structure in which the axis of pseudospin rotation twists as a function of $y$.
In this case, the second term in Eq.~(\ref{eq:spin-charge-final}) is expected to result in extra net charge on top of the dipole, while the spin density components will also display more complex patterns.

We use a rotation operator that results in a similar domain structure as before, but, instead of implementing a uniaxial pseudospin rotation along $\hat{\vec{y}}$, we employ a $y$-dependent ``twisting'' rotation axis $( \sin [2\pi y/L_y], \cos[2\pi y/L_y], 0 )$, see App.~\ref{app:twist} for the details.
The resulting pseudospin texture is shown in Fig.~\ref{fig:num2}a.
We plot the calculated $z$-component of the pseudospin density in color scale, and use pink arrows to indicate the local direction of $\langle \overline{\vec{\tau}}(\vec{r})\rangle$; the calculated in-plane components $\langle \overline{\tau}^{x,y}(\vec{r})\rangle$ are shown in the two leftmost panels of Fig.~\ref{fig:num2}c.
The operation $U_R$ that we applied indeed results in a domain wall with an in-plane twist.

In Fig.~\ref{fig:num2}b we show the calculated change in charge density $\langle \delta \overline{\rho}(\vec{r}) \rangle = \langle \overline{\rho}(\vec{r})\rangle - \rho_0$, and in the two rightmost panels in Fig.~\ref{fig:num2}c present the in-plane spin densities.
We see that both $\langle \delta \overline{\rho}(\vec{r}) \rangle$ and $\langle \overline{s}^{x,y}(\vec{r})\rangle$ no longer have a clear dipole structure around the domain wall (as in the case of a uniaxial rotation studied above), but rather show a net charge and ($y$-dependent) spin density localized at the domain wall.
Equation (\ref{eq:spin-charge-final}) predicts an extra (integrated) charge of 1 to localize at the domain wall for the ``$2\pi$-twist'' we created; numerically we find an excess charge of 0.996, which agrees well with the expectation.
More generally, we note that a $2n\pi$-twist, where the rotation axis is $( \sin [2n\pi y/L_y], \cos[2n\pi y/L_y], 0 )$, will result in an extra charge of $n$ electrons bound at the domain wall~\cite{Jungwirth01}.

To compare the numerical results again with our analytic expressions, in Fig.~\ref{fig:num2}d we show  both line cuts of Figs.~\ref{fig:num2}a,b at $y=25\, \ell$ (dots) and the analytically calculated $\langle \overline{\tau}^{z}(\vec{r})\rangle$ (red, left axis) and $\langle \delta \overline{\rho}(\vec{r}) \rangle$ (black, right axis) along the same line.
Figs.~\ref{fig:num2}e--g show line cuts of the in-plane spin densities $\langle \overline{s}^{x}(\vec{r})\rangle$ (green) and $\langle \overline{s}^{y}(\vec{r})\rangle$ (blue), along $y = 25\,\ell$ (e), $y = 5\,\ell$ (f), and $x=0$ (g); these lines are indicated with red dotted lines in Fig.~\ref{fig:num2}c.
We again see that all calculated quantities show good agreement between the numerical results and our analytic expressions.
Based on Eq.~(\ref{eq:spin-charge-final}), the charge density still has a dipole-like contribution that is of the same magnitude as in the untwisted domain wall.
This contribution, however, is dominated by the second term in (\ref{eq:spin-charge-final}), the Pontryagin index density, that results in the localization of excess charges at the domain wall.
The in-plane spin densities also no longer show a dipole structure, but rather exhibit a localization of excess in-plane spin close to the domain wall, similar to the charge density.

\section{Conclusion and discussion}\label{sec:Discussion}

We have thus characterized, both analytically and numerically, the charge and spin textures that arise from smooth domain walls in Ising quantum Hall ferromagnets, where the two polarizations (denoted `pseudospin' up and down) correspond to states with opposite physical spin and different Landau level index, $n\uparrow$ and $m\downarrow$.
For convenience we worked in the Landau gauge throughout, but similar considerations can be applied in other gauges (such as the symmetric gauge) to study domain walls with other geometries such as domain wall loops.

The charge and spin textures that we studied may have important implications for the energetics and dynamics of domain walls in Ising quantum Hall systems.
For example, the charge dipole that unavoidably accompanies even neutral domains may play role in the interaction between domains, as well as their interaction with external electrostatic potentials.
We expect our results to be of particular relevance for the phenomenon of ``nuclear electric resonance,'' where radio frequency modulations of electrostatic gates are used to induce nuclear spin resonance via the nuclear spins' coupling to electron spins in such domain walls~\cite{Miyamoto16,Korkusinski17}.
In this context, both the electrostatics of the domain walls and their (physical) spin structures are expected to play important roles.

Throughout this work, we focused on the integer quantum Hall effect regime. 
Ising-type domain walls in the fractional quantum Hall effect regime have also been the subject of intense experimental interest.
Extending our results to the fractional quantum Hall regime and applying them to modeling of nuclear electric resonance are interesting directions for future work.

\begin{acknowledgments}
We are grateful for helpful discussions with E. Berg, M. Levin, I. Neder, and M. Zaletel.
The Center for Quantum Devices is funded by the Danish National Research Foundation.
This work was supported by the European Research Council (ERC) under the European Union Horizon 2020 Research and Innovation Programme, Grant Agreement No. 678862, the Villum Foundation, and the Research Council of Norway through its Centers of Excellence funding scheme, project number 262633, QuSpin.\\
\end{acknowledgments}

\appendix

\section{Explicit form of projected operators and some useful identities} 
\label{appendix_explicit_definitions}

Here we give the explicit forms of the many-particle wave function and the projected operators that we defined in the main text.
We construct the homogeneously polarized initial state using the single-particle states
\begin{align}
    \ket{\psi_k} =
    \begin{pmatrix}
        \cos (\theta/2) \ket{n,k}\\
        e^{i\phi}\sin (\theta/2) \ket{m,k}
    \end{pmatrix},
    \label{eq:sp-orbitals_general}
\end{align}
where the explicit forms of the LL wave functions $\ket{n,k}$ are given in Eq.~(\ref{eq:llwavefunc}), assuming the Landau gauge.
Each single-particle state is thus constructed with a pseudospin polarization along direction $(\theta,\phi)$ of the Bloch sphere.

Using these single-particle states, we write the homogeneous many-body state as a Slater determinant,
\begin{align}
|\Psi_0 \rangle=\frac{1}{\sqrt{N!}}\,\det \big\{ |{\psi^{(j)}_{k}}\rangle\big\},
\end{align}
where the index $j=1,\dots,N$ labels the $N$ particles and $k = 2\pi l/L_{y}$, where $l$ is an integer in the range $-(N-1)/2\leq l \leq (N-1)/2$ (using odd $N$).
The many-particle rotation operator $U_R$ defined in Eq.~(\ref{eq:R}) is formally a tensor product of single-particle rotation operators, $U_R = u_R^{(1)} \otimes u_R^{(2)} \otimes \cdots \otimes u_R^{(N)}$.
The expectation value of a many-body operator $ A = \sum_j a^{(j)}$ reduces to:
\begin{equation}
\langle A \rangle = \langle {\Psi_0|U_R^\dagger\, A\, U_R|\Psi_0} \rangle = \sum_k \langle {\psi_k|u_R^\dagger\, a\, u_R|\psi_k} \rangle.
\end{equation}

Using the convention for the Fourier transform
\begin{eqnarray}
f(\vec{r})&=&\sum_{p_{y}} \int_{-\infty}^{\infty} \!\! dp_{x}\,f_{\vec{p}}\,e^{-i \vec{p}\cdot \vec{r}}, \\
f_{\vec{p}}&=&\frac{1}{2\pi L_{y}} \int_{0}^{L_{y}} \!\! dy \int_{-\infty}^{\infty} \!\! dx\,f(\vec{r})e^{i \vec{p}\cdot \vec{r}},
\end{eqnarray}
we straightforwardly find the Fourier transformed projected pseudospin density operators:
\begin{widetext}
\begin{align}
\bar\tau_{\vec{q}}^x = {} & {} \frac{1}{2} \frac{1}{2\pi L_y} \sum_k \big( w^{n,n}_{k,{\vec{q}}} + w^{m,m}_{k,{\vec{q}}} \big)
\left( \begin{array}{cc}
0 & \ket{n,k}\bra{m,k+q_y} \\
\ket{m,k}\bra{n,k+q_y} & 0 
\end{array}\right),\label{eq:taux_expl} \\
\bar\tau_{\vec{q}}^y = {} & {} \frac{1}{2} \frac{1}{2\pi L_y} \sum_k \big( w^{n,n}_{k,{\vec{q}}} + w^{m,m}_{k,{\vec{q}}} \big)
\left( \begin{array}{cc}
0 & -i\ket{n,k}\bra{m,k+q_y} \\
i\ket{m,k}\bra{n,k+q_y} & 0 
\end{array}\right), \\
\bar\tau_{\vec{q}}^z = {} & {} \frac{1}{2\pi L_y} \sum_k \left( \begin{array}{cc}
w^{n,n}_{k,{\vec{q}}} \ket{n,k}\bra{n,k+q_y} & 0 \\
0 & -w^{m,m}_{k,{\vec{q}}} \ket{m,k}\bra{m,k+q_y}
\end{array}\right),\label{eq:tauz_expl}
\end{align}
where $w^{n,m}_{k,\vec{q}}$ stands for a Fourier-transformed overlap of LL wave functions $\psi_{n,k}(\vec{r})$ and $\psi_{m,k+q_y}(\vec{r})$ [see Eq.~(\ref{eq:llwavefunc})]:
\begin{align}
w^{n,m}_{k,\vec{q}} \equiv \int_0^{L_y} \!\! dy\int_{-\infty}^\infty \!\! dx\,
\psi_{n,k}({\vec{r}})^*\psi_{m,k+q_y}({\vec{r}}) e^{i{\vec{q}}\cdot{\vec{r}}}.
\end{align}
Similarly we find explicit expressions for the projected physical spin density operators,
\begin{align}
\overline{\sigma}^x_{\vec{q}} = 
{} & {} \frac{1}{2\pi L_y} \sum_k \left( \begin{array}{cc}
0 & w^{n,m}_{k,{\vec{q}}} \ket{n,k}\bra{m,k+q_y} \\
w^{m,n}_{k,{\vec{q}}} \ket{m,k}\bra{n,k+q_y} & 0 
\end{array}\right),\\
\overline{\sigma}^y_{\vec{q}} = 
{} & {} \frac{1}{2\pi L_y} \sum_k \left( \begin{array}{cc}
0 & -i w^{n,m}_{k,{\vec{q}}} \ket{n,k}\bra{m,k+q_y} \\
i w^{m,n}_{k,{\vec{q}}} \ket{m,k}\bra{n,k+q_y} & 0 
\end{array}\right), \\
\overline{\sigma}^z_{\vec{q}} = {} & {} \frac{1}{2\pi L_y} \sum_k \left( \begin{array}{cc}
w^{n,n}_{k,{\vec{q}}} \ket{n,k}\bra{n,k+q_y} & 0 \\
0 & -w^{m,m}_{k,{\vec{q}}} \ket{m,k}\bra{m,k+q_y}
\end{array}\right)
= \bar\tau^z_{\vec{q}},
\end{align}
and the projected charge density operator,
\begin{align}
\overline{\rho_{\vec{q}}} = \bar\tau_{\vec{q}}^0 = {} & {} \frac{1}{2\pi L_y} \sum_k \left( \begin{array}{cc}
w^{n,n}_{k,{\vec{q}}} \ket{n,k}\bra{n,k+q_y} & 0 \\
0 & w^{m,m}_{k,{\vec{q}}} \ket{m,k}\bra{m,k+q_y}
\end{array}\right).
\end{align}

\section{Creating a pseudospin texture}

Using the definitions of the Fourier transforms presented in App.~\ref{appendix_explicit_definitions},  we express the operator $\overline{O}$ in Eq.~(\ref{eq:R}) as:
\begin{equation}
\overline{{O}} = 2 \pi L_{y} \sum_{\eta=x,y,z}\sum_{q_{y}} \int_{-\infty}^\infty \!\! dq_{x}\, \frac{1}{2}\Omega^{\eta}_{\vec{q}}\overline{\tau}^{\eta}_{-\vec{q}},
\end{equation}
where
\begin{eqnarray}
\Omega^{\eta}_{\vec{q}}&=&\frac{1}{2\pi L_{y}} \int_{0}^{L_{y}} \! \! dy \int_{-\infty}^{\infty}\!\! dx\,\Omega^{\eta}(\vec{r})\, e^{i \vec{q}\cdot \vec{r}}.
\end{eqnarray}

\subsection{Homogeneous domain wall}\label{app:homodw}

For the case of the uniform domain wall as considered in the main text, we used $\vec{\Omega}(\vec{r}) =(0,f(x),0 )\equiv \Omega^{y}(x) \vec{\hat{y}}$.
Then the explicit form of $\overline{{O}}$ follows as
\begin{equation}
\overline{{O}}= 
    \frac{1}{4} 
    \sum_{k} 
    \int^{\infty}_{-\infty} \!\! dx \, f(x)\left(|\phi_{n,k}(x)|^{2}+ |\phi_{m,k}(x)|^{2} \right)
    \begin{pmatrix}
        0 & -i\ketbra{n,k}{m,k}\\
        i\ketbra{m,k}{n,k} & 0
    \end{pmatrix},
\label{eq:explicit_sg_O_dw_perfect_exemplary} 
\end{equation}
where
\begin{align}
\phi_{n,k}(x)  = {} & {} \frac{1}{\sqrt{2^n n! \ell \sqrt \pi }}
\, H_n \! \left( \frac{x - k\ell^2}{\ell} \right) e^{ - \frac{ (x-k\ell^2)^2 }{ 2\ell^2 }}, \label{eq:x_llwavefunc}
\end{align}
is the $x$-dependent part of the wave function $\psi_{n,k}(\vec{r})$.
Owing to the fact that $\Omega^{y}(x)$ is only a function of $x$, the operator $\overline{{O}}$ is diagonal in wave number $k$.
In this simple case (see below for a counterexample), the function $f(x)$ coincides with the actual rotation function that produces the resulting pseudospin structure, i.e., the vector field
\begin{align}
\vec{n}(\vec{r}) = R[ f(x) \vec{\hat{y}} ] \, ( 0, 0, 1 )^{\rm T}
\end{align}
is parallel to the resulting $\langle \overline{\vec{\tau}}(\vec{r} ) \rangle$ everywhere.
The three-dimensional rotation operator $R[ \vec{u} ]$ is defined to produce a rotation over an angle $\phi = |\vec{u}|$ along the axis $\hat{\vec s} = \vec{u}/u$, and reads explicitly as:
\begin{align}
R[ \vec{u} ] = \left( \begin{array}{ccc}
\cos \phi + (\hat{s}^x)^2(1-\cos\phi) & \hat{s}^x\hat{s}^y(1-\cos\phi) - \hat{s}^z \sin\phi & \hat{s}^x\hat{s}^z(1-\cos\phi) + \hat{s}^y\sin\phi \\
\hat{s}^y\hat{s}^x(1-\cos\phi) + \hat{s}^z \sin\phi & \cos \phi + (\hat{s}^y)^2(1-\cos\phi) & \hat{s}^y\hat{s}^z(1-\cos\phi) - \hat{s}^x\sin\phi \\
\hat{s}^z\hat{s}^x(1-\cos\phi) - \hat{s}^y \sin\phi & \hat{s}^z\hat{s}^y(1-\cos\phi) + \hat{s}^x \sin\phi & \cos\phi + (\hat s^z)^2(1-\cos\phi)
\end{array}\right).\label{eq:3drot}
\end{align}

\subsection{Domain wall with a twist}\label{app:twist}

In Sec.~\ref{sec:non-uni} we studied twisted pseudospin textures, where the in-plane components of the pseudospin density wind $p$ times along the domain wall (focusing on the case $p = 1$ for demonstration).
We aim to create a texture described by the rotation function $f(x)(\sin[2\pi p y/L_y],\cos[2\pi p y/L_y],0)$.
Starting from a uniform state, we apply the pseudospin rotation in Eq.~(\ref{eq:R}) of the main text, with $\overline{O}(p)$ given by:
\begin{align}
\label{eq:ObarRenorm}
\overline{O}(p) = \frac{1}{4} \sum_k 
\left( \begin{array}{cc}
0 & -i(\tilde f^n_{k,p} + \tilde f^m_{k,p})  \ket{n,k}\bra{m,k+2\pi p/L_y} \\
i(\tilde f^n_{k,-p} + \tilde f^m_{k,-p}) \ket{m,k}\bra{n,k-2 \pi p/L_y} & 0 
\end{array}\right).
\end{align}
Importantly, $\overline{O}(p)$ is {\it not} diagonal in $k$ space, and its norm is slightly suppressed due to the non-unit overlap between shifted orbitals $\phi_{n,k}(x)$ and $\phi_{n,k + 2\pi p/L_y}(x)$.
To achieve the desired rotation, we therefore employ the renormalized weight factors $\tilde f^n_{k,p}$ in Eq.~(\ref{eq:ObarRenorm}):
\begin{align}
\tilde f^n_{k,p} = \frac{ \int^{\infty}_{-\infty} dx\, f(x) [\phi_{n,k}(x)]^{*}\phi_{n,k+ 2\pi p/L_y}(x)}{\int^{\infty}_{-\infty} dx\, [\phi_{n,k}(x)]^{*}\phi_{n,k+ 2\pi p/L_y}(x)}.
\end{align}
\end{widetext}


\begin{thebibliography}{51}
	\expandafter\ifx\csname natexlab\endcsname\relax\def\natexlab#1{#1}\fi
	\expandafter\ifx\csname bibnamefont\endcsname\relax
	\def\bibnamefont#1{#1}\fi
	\expandafter\ifx\csname bibfnamefont\endcsname\relax
	\def\bibfnamefont#1{#1}\fi
	\expandafter\ifx\csname citenamefont\endcsname\relax
	\def\citenamefont#1{#1}\fi
	\expandafter\ifx\csname url\endcsname\relax
	\def\url#1{\texttt{#1}}\fi
	\expandafter\ifx\csname urlprefix\endcsname\relax\def\urlprefix{URL }\fi
	\providecommand{\bibinfo}[2]{#2}
	\providecommand{\eprint}[2][]{\url{#2}}
	
	\bibitem[{\citenamefont{Girvin}(2000)}]{Girvin00}
	\bibinfo{author}{\bibfnamefont{S.~M.} \bibnamefont{Girvin}},
	\bibinfo{journal}{Physics Today} \textbf{\bibinfo{volume}{53}},
	\bibinfo{pages}{39} (\bibinfo{year}{2000}).
	
	\bibitem[{\citenamefont{Sondhi et~al.}(1993)\citenamefont{Sondhi, Karlhede,
			Kivelson, and Rezayi}}]{Sondhi93}
	\bibinfo{author}{\bibfnamefont{S.~L.} \bibnamefont{Sondhi}},
	\bibinfo{author}{\bibfnamefont{A.}~\bibnamefont{Karlhede}},
	\bibinfo{author}{\bibfnamefont{S.~A.} \bibnamefont{Kivelson}},
	\bibnamefont{and} \bibinfo{author}{\bibfnamefont{E.~H.}
		\bibnamefont{Rezayi}}, \bibinfo{journal}{Phys. Rev. B}
	\textbf{\bibinfo{volume}{47}}, \bibinfo{pages}{16419} (\bibinfo{year}{1993}).
	
	\bibitem[{\citenamefont{Wen and Zee}(1992)}]{Wen92c}
	\bibinfo{author}{\bibfnamefont{X.-G.} \bibnamefont{Wen}} \bibnamefont{and}
	\bibinfo{author}{\bibfnamefont{A.}~\bibnamefont{Zee}},
	\bibinfo{journal}{Phys. Rev. Lett.} \textbf{\bibinfo{volume}{69}},
	\bibinfo{pages}{1811} (\bibinfo{year}{1992}).
	
	\bibitem[{\citenamefont{Yang et~al.}(1994)\citenamefont{Yang, Moon, Zheng,
			MacDonald, Girvin, Yoshioka, and Zhang}}]{Yang94}
	\bibinfo{author}{\bibfnamefont{K.}~\bibnamefont{Yang}},
	\bibinfo{author}{\bibfnamefont{K.}~\bibnamefont{Moon}},
	\bibinfo{author}{\bibfnamefont{L.}~\bibnamefont{Zheng}},
	\bibinfo{author}{\bibfnamefont{A.~H.} \bibnamefont{MacDonald}},
	\bibinfo{author}{\bibfnamefont{S.~M.} \bibnamefont{Girvin}},
	\bibinfo{author}{\bibfnamefont{D.}~\bibnamefont{Yoshioka}}, \bibnamefont{and}
	\bibinfo{author}{\bibfnamefont{S.-C.} \bibnamefont{Zhang}},
	\bibinfo{journal}{Phys. Rev. Lett.} \textbf{\bibinfo{volume}{72}},
	\bibinfo{pages}{732} (\bibinfo{year}{1994}).
	
	\bibitem[{\citenamefont{Smet et~al.}(2001)\citenamefont{Smet, Deutschmann,
			Wegscheider, Abstreiter, and von Klitzing}}]{Smet01}
	\bibinfo{author}{\bibfnamefont{J.~H.} \bibnamefont{Smet}},
	\bibinfo{author}{\bibfnamefont{R.~A.} \bibnamefont{Deutschmann}},
	\bibinfo{author}{\bibfnamefont{W.}~\bibnamefont{Wegscheider}},
	\bibinfo{author}{\bibfnamefont{G.}~\bibnamefont{Abstreiter}},
	\bibnamefont{and} \bibinfo{author}{\bibfnamefont{K.}~\bibnamefont{von
			Klitzing}}, \bibinfo{journal}{Phys.~Rev.~Lett.}
	\textbf{\bibinfo{volume}{86}}, \bibinfo{pages}{2412} (\bibinfo{year}{2001}).
	
	\bibitem[{\citenamefont{Muraki et~al.}(2001)\citenamefont{Muraki, Saku, and
			Hirayama}}]{Muraki01}
	\bibinfo{author}{\bibfnamefont{K.}~\bibnamefont{Muraki}},
	\bibinfo{author}{\bibfnamefont{T.}~\bibnamefont{Saku}}, \bibnamefont{and}
	\bibinfo{author}{\bibfnamefont{Y.}~\bibnamefont{Hirayama}},
	\bibinfo{journal}{Phys. Rev. Lett.} \textbf{\bibinfo{volume}{87}},
	\bibinfo{pages}{196801} (\bibinfo{year}{2001}).
	
	\bibitem[{\citenamefont{Jaroszy\ifmmode~\acute{n}\else \'{n}\fi{}ski
			et~al.}(2002)\citenamefont{Jaroszy\ifmmode~\acute{n}\else \'{n}\fi{}ski,
			Andrearczyk, Karczewski, Wr\'obel, Wojtowicz, Papis,
			Kami\ifmmode~\acute{n}\else \'{n}\fi{}ska, Piotrowska,
			Popovi\ifmmode~\acute{c}\else \'{c}\fi{}, and Dietl}}]{Jaroszynski02}
	\bibinfo{author}{\bibfnamefont{J.}~\bibnamefont{Jaroszy\ifmmode~\acute{n}\else
			\'{n}\fi{}ski}},
	\bibinfo{author}{\bibfnamefont{T.}~\bibnamefont{Andrearczyk}},
	\bibinfo{author}{\bibfnamefont{G.}~\bibnamefont{Karczewski}},
	\bibinfo{author}{\bibfnamefont{J.}~\bibnamefont{Wr\'obel}},
	\bibinfo{author}{\bibfnamefont{T.}~\bibnamefont{Wojtowicz}},
	\bibinfo{author}{\bibfnamefont{E.}~\bibnamefont{Papis}},
	\bibinfo{author}{\bibfnamefont{E.}~\bibnamefont{Kami\ifmmode~\acute{n}\else
			\'{n}\fi{}ska}},
	\bibinfo{author}{\bibfnamefont{A.}~\bibnamefont{Piotrowska}},
	\bibinfo{author}{\bibfnamefont{D.}~\bibnamefont{Popovi\ifmmode~\acute{c}\else
			\'{c}\fi{}}}, \bibnamefont{and}
	\bibinfo{author}{\bibfnamefont{T.}~\bibnamefont{Dietl}},
	\bibinfo{journal}{Phys. Rev. Lett.} \textbf{\bibinfo{volume}{89}},
	\bibinfo{pages}{266802} (\bibinfo{year}{2002}).
	
	\bibitem[{\citenamefont{Lok et~al.}(2004)\citenamefont{Lok, Lynass, Dietsche,
			von Klitzing, and Hauser}}]{Lok04}
	\bibinfo{author}{\bibfnamefont{J.~G.~S.} \bibnamefont{Lok}},
	\bibinfo{author}{\bibfnamefont{M.}~\bibnamefont{Lynass}},
	\bibinfo{author}{\bibfnamefont{W.}~\bibnamefont{Dietsche}},
	\bibinfo{author}{\bibfnamefont{K.}~\bibnamefont{von Klitzing}},
	\bibnamefont{and} \bibinfo{author}{\bibfnamefont{M.}~\bibnamefont{Hauser}},
	\bibinfo{journal}{Physica E} \textbf{\bibinfo{volume}{22}},
	\bibinfo{pages}{94} (\bibinfo{year}{2004}).
	
	\bibitem[{\citenamefont{Nomura and MacDonald}(2006)}]{Nomura06}
	\bibinfo{author}{\bibfnamefont{K.}~\bibnamefont{Nomura}} \bibnamefont{and}
	\bibinfo{author}{\bibfnamefont{A.~H.} \bibnamefont{MacDonald}},
	\bibinfo{journal}{Phys. Rev. Lett.} \textbf{\bibinfo{volume}{96}},
	\bibinfo{pages}{256602} (\bibinfo{year}{2006}).
	
	\bibitem[{\citenamefont{Feldman et~al.}(2009)\citenamefont{Feldman, Martin, and
			Yacoby}}]{Feldman09}
	\bibinfo{author}{\bibfnamefont{B.~E.} \bibnamefont{Feldman}},
	\bibinfo{author}{\bibfnamefont{J.}~\bibnamefont{Martin}}, \bibnamefont{and}
	\bibinfo{author}{\bibfnamefont{A.}~\bibnamefont{Yacoby}},
	\bibinfo{journal}{Nature Physics} \textbf{\bibinfo{volume}{5}},
	\bibinfo{pages}{889} (\bibinfo{year}{2009}).
	
	\bibitem[{\citenamefont{Weitz et~al.}(2010)\citenamefont{Weitz, Allen, Feldman,
			Martin, and Yacoby}}]{Weitz2010}
	\bibinfo{author}{\bibfnamefont{R.~T.} \bibnamefont{Weitz}},
	\bibinfo{author}{\bibfnamefont{M.~T.} \bibnamefont{Allen}},
	\bibinfo{author}{\bibfnamefont{B.~E.} \bibnamefont{Feldman}},
	\bibinfo{author}{\bibfnamefont{J.}~\bibnamefont{Martin}}, \bibnamefont{and}
	\bibinfo{author}{\bibfnamefont{A.}~\bibnamefont{Yacoby}},
	\bibinfo{journal}{Science} \textbf{\bibinfo{volume}{330}},
	\bibinfo{pages}{812} (\bibinfo{year}{2010}).
	
	\bibitem[{\citenamefont{C\^ot\'e et~al.}(2010)\citenamefont{C\^ot\'e, Luo,
			Petrov, Barlas, and MacDonald}}]{Cote10}
	\bibinfo{author}{\bibfnamefont{R.}~\bibnamefont{C\^ot\'e}},
	\bibinfo{author}{\bibfnamefont{W.}~\bibnamefont{Luo}},
	\bibinfo{author}{\bibfnamefont{B.}~\bibnamefont{Petrov}},
	\bibinfo{author}{\bibfnamefont{Y.}~\bibnamefont{Barlas}}, \bibnamefont{and}
	\bibinfo{author}{\bibfnamefont{A.~H.} \bibnamefont{MacDonald}},
	\bibinfo{journal}{Phys. Rev. B} \textbf{\bibinfo{volume}{82}},
	\bibinfo{pages}{245307} (\bibinfo{year}{2010}).
	
	\bibitem[{\citenamefont{Dean et~al.}(2011)\citenamefont{Dean, Young,
			Cadden-Zimansky, Wang, Ren, Watanabe, Taniguchi, Kim, Hone, and
			Shepard}}]{Dean11}
	\bibinfo{author}{\bibfnamefont{C.~R.} \bibnamefont{Dean}},
	\bibinfo{author}{\bibfnamefont{A.~F.} \bibnamefont{Young}},
	\bibinfo{author}{\bibfnamefont{P.}~\bibnamefont{Cadden-Zimansky}},
	\bibinfo{author}{\bibfnamefont{L.}~\bibnamefont{Wang}},
	\bibinfo{author}{\bibfnamefont{H.}~\bibnamefont{Ren}},
	\bibinfo{author}{\bibfnamefont{K.}~\bibnamefont{Watanabe}},
	\bibinfo{author}{\bibfnamefont{T.}~\bibnamefont{Taniguchi}},
	\bibinfo{author}{\bibfnamefont{P.}~\bibnamefont{Kim}},
	\bibinfo{author}{\bibfnamefont{J.}~\bibnamefont{Hone}}, \bibnamefont{and}
	\bibinfo{author}{\bibfnamefont{K.~L.} \bibnamefont{Shepard}},
	\bibinfo{journal}{Nature Physics} \textbf{\bibinfo{volume}{7}},
	\bibinfo{pages}{693} (\bibinfo{year}{2011}).
	
	\bibitem[{\citenamefont{Young et~al.}(2012)\citenamefont{Young, Dean, Wang,
			Ren, Cadden-Zimansky, Watanabe, Taniguchi, Hone, Shepard, and Kim}}]{Young12}
	\bibinfo{author}{\bibfnamefont{A.~F.} \bibnamefont{Young}},
	\bibinfo{author}{\bibfnamefont{C.~R.} \bibnamefont{Dean}},
	\bibinfo{author}{\bibfnamefont{L.}~\bibnamefont{Wang}},
	\bibinfo{author}{\bibfnamefont{H.}~\bibnamefont{Ren}},
	\bibinfo{author}{\bibfnamefont{P.}~\bibnamefont{Cadden-Zimansky}},
	\bibinfo{author}{\bibfnamefont{K.}~\bibnamefont{Watanabe}},
	\bibinfo{author}{\bibfnamefont{T.}~\bibnamefont{Taniguchi}},
	\bibinfo{author}{\bibfnamefont{J.}~\bibnamefont{Hone}},
	\bibinfo{author}{\bibfnamefont{K.~L.} \bibnamefont{Shepard}},
	\bibnamefont{and} \bibinfo{author}{\bibfnamefont{P.}~\bibnamefont{Kim}},
	\bibinfo{journal}{Nat Phys} \textbf{\bibinfo{volume}{8}},
	\bibinfo{pages}{550} (\bibinfo{year}{2012}).
	
	\bibitem[{\citenamefont{Feldman et~al.}(2016)\citenamefont{Feldman, Randeria,
			Gyenis, We, Ji, Cava, MacDonald, and Yazdani}}]{Feldman16}
	\bibinfo{author}{\bibfnamefont{B.~E.} \bibnamefont{Feldman}},
	\bibinfo{author}{\bibfnamefont{M.}~\bibnamefont{Randeria}},
	\bibinfo{author}{\bibfnamefont{A.}~\bibnamefont{Gyenis}},
	\bibinfo{author}{\bibfnamefont{F.}~\bibnamefont{We}},
	\bibinfo{author}{\bibfnamefont{H.}~\bibnamefont{Ji}},
	\bibinfo{author}{\bibfnamefont{R.~J.} \bibnamefont{Cava}},
	\bibinfo{author}{\bibfnamefont{A.~H.} \bibnamefont{MacDonald}},
	\bibnamefont{and} \bibinfo{author}{\bibfnamefont{A.}~\bibnamefont{Yazdani}},
	\bibinfo{journal}{Science} \textbf{\bibinfo{volume}{354}},
	\bibinfo{pages}{316} (\bibinfo{year}{2016}).
	
	\bibitem[{\citenamefont{Sodemann et~al.}(2017)\citenamefont{Sodemann, Zhu, and
			Fu}}]{Sodemann17}
	\bibinfo{author}{\bibfnamefont{I.}~\bibnamefont{Sodemann}},
	\bibinfo{author}{\bibfnamefont{Z.}~\bibnamefont{Zhu}}, \bibnamefont{and}
	\bibinfo{author}{\bibfnamefont{L.}~\bibnamefont{Fu}}, \bibinfo{journal}{Phys.
		Rev. X} \textbf{\bibinfo{volume}{7}}, \bibinfo{pages}{041068}
	(\bibinfo{year}{2017}).
	
	\bibitem[{\citenamefont{Piazza et~al.}(1999)\citenamefont{Piazza, Pellegrini,
			Beltram, Wegscheider, Jungwirth, and MacDonald}}]{Piazza99}
	\bibinfo{author}{\bibfnamefont{V.}~\bibnamefont{Piazza}},
	\bibinfo{author}{\bibfnamefont{V.}~\bibnamefont{Pellegrini}},
	\bibinfo{author}{\bibfnamefont{F.}~\bibnamefont{Beltram}},
	\bibinfo{author}{\bibfnamefont{W.}~\bibnamefont{Wegscheider}},
	\bibinfo{author}{\bibfnamefont{T.}~\bibnamefont{Jungwirth}},
	\bibnamefont{and} \bibinfo{author}{\bibfnamefont{A.~H.}
		\bibnamefont{MacDonald}}, \bibinfo{journal}{Nature}
	\textbf{\bibinfo{volume}{402}}, \bibinfo{pages}{638} (\bibinfo{year}{1999}).
	
	\bibitem[{\citenamefont{De~Poortere et~al.}(2000)\citenamefont{De~Poortere,
			Tutuc, Papadakis, and Shayegan}}]{DePoortere00}
	\bibinfo{author}{\bibfnamefont{E.~P.} \bibnamefont{De~Poortere}},
	\bibinfo{author}{\bibfnamefont{E.}~\bibnamefont{Tutuc}},
	\bibinfo{author}{\bibfnamefont{S.~J.} \bibnamefont{Papadakis}},
	\bibnamefont{and} \bibinfo{author}{\bibfnamefont{M.}~\bibnamefont{Shayegan}},
	\bibinfo{journal}{Science} \textbf{\bibinfo{volume}{290}},
	\bibinfo{pages}{1546} (\bibinfo{year}{2000}).
	
	\bibitem[{\citenamefont{Jungwirth and
			MacDonald}(2001{\natexlab{a}})}]{Jungwirth01}
	\bibinfo{author}{\bibfnamefont{T.}~\bibnamefont{Jungwirth}} \bibnamefont{and}
	\bibinfo{author}{\bibfnamefont{A.~H.} \bibnamefont{MacDonald}},
	\bibinfo{journal}{Phys. Rev. Lett.} \textbf{\bibinfo{volume}{87}},
	\bibinfo{pages}{216801} (\bibinfo{year}{2001}{\natexlab{a}}).
	
	\bibitem[{\citenamefont{Brey and Tejedor}(2002)}]{PhysRevB.66.041308}
	\bibinfo{author}{\bibfnamefont{L.}~\bibnamefont{Brey}} \bibnamefont{and}
	\bibinfo{author}{\bibfnamefont{C.}~\bibnamefont{Tejedor}},
	\bibinfo{journal}{Phys. Rev. B} \textbf{\bibinfo{volume}{66}},
	\bibinfo{pages}{041308} (\bibinfo{year}{2002}).
	
	\bibitem[{\citenamefont{Kumada et~al.}(2008)\citenamefont{Kumada, Kamada,
			Miyashita, Hirayama, and Fujisawa}}]{Kumada08}
	\bibinfo{author}{\bibfnamefont{N.}~\bibnamefont{Kumada}},
	\bibinfo{author}{\bibfnamefont{T.}~\bibnamefont{Kamada}},
	\bibinfo{author}{\bibfnamefont{S.}~\bibnamefont{Miyashita}},
	\bibinfo{author}{\bibfnamefont{Y.}~\bibnamefont{Hirayama}}, \bibnamefont{and}
	\bibinfo{author}{\bibfnamefont{T.}~\bibnamefont{Fujisawa}},
	\bibinfo{journal}{Phys. Rev. Lett.} \textbf{\bibinfo{volume}{101}},
	\bibinfo{pages}{137602} (\bibinfo{year}{2008}).
	
	\bibitem[{\citenamefont{Yusa et~al.}(2004)\citenamefont{Yusa, Hashimoto,
			Muraki, Saku, and Hirayama}}]{Yusa04}
	\bibinfo{author}{\bibfnamefont{G.}~\bibnamefont{Yusa}},
	\bibinfo{author}{\bibfnamefont{K.}~\bibnamefont{Hashimoto}},
	\bibinfo{author}{\bibfnamefont{K.}~\bibnamefont{Muraki}},
	\bibinfo{author}{\bibfnamefont{T.}~\bibnamefont{Saku}}, \bibnamefont{and}
	\bibinfo{author}{\bibfnamefont{Y.}~\bibnamefont{Hirayama}},
	\bibinfo{journal}{Phys. Rev. B} \textbf{\bibinfo{volume}{69}},
	\bibinfo{pages}{161302} (\bibinfo{year}{2004}).
	
	\bibitem[{\citenamefont{Liu et~al.}(2010)\citenamefont{Liu, Yang, Mishima,
			Santos, and Hirayama}}]{Liu10}
	\bibinfo{author}{\bibfnamefont{H.~W.} \bibnamefont{Liu}},
	\bibinfo{author}{\bibfnamefont{K.~F.} \bibnamefont{Yang}},
	\bibinfo{author}{\bibfnamefont{T.~D.} \bibnamefont{Mishima}},
	\bibinfo{author}{\bibfnamefont{M.~B.} \bibnamefont{Santos}},
	\bibnamefont{and} \bibinfo{author}{\bibfnamefont{Y.}~\bibnamefont{Hirayama}},
	\bibinfo{journal}{Phys. Rev. B} \textbf{\bibinfo{volume}{82}},
	\bibinfo{pages}{241304} (\bibinfo{year}{2010}).
	
	\bibitem[{\citenamefont{Lu et~al.}(2017)\citenamefont{Lu, Tracy, Laroche,
			Huang, Chuang, Su, Li, and Liu}}]{Lu17}
	\bibinfo{author}{\bibfnamefont{T.~M.} \bibnamefont{Lu}},
	\bibinfo{author}{\bibfnamefont{L.~A.} \bibnamefont{Tracy}},
	\bibinfo{author}{\bibfnamefont{D.}~\bibnamefont{Laroche}},
	\bibinfo{author}{\bibfnamefont{S.-H.} \bibnamefont{Huang}},
	\bibinfo{author}{\bibfnamefont{Y.}~\bibnamefont{Chuang}},
	\bibinfo{author}{\bibfnamefont{Y.-H.} \bibnamefont{Su}},
	\bibinfo{author}{\bibfnamefont{J.-Y.} \bibnamefont{Li}}, \bibnamefont{and}
	\bibinfo{author}{\bibfnamefont{C.~W.} \bibnamefont{Liu}},
	\bibinfo{journal}{Scientific Reports} \textbf{\bibinfo{volume}{7}},
	\bibinfo{pages}{2468} (\bibinfo{year}{2017}).
	
	\bibitem[{\citenamefont{Korkusinski et~al.}(2017)\citenamefont{Korkusinski,
			Hawrylak, Liu, and Hirayama}}]{Korkusinski17}
	\bibinfo{author}{\bibfnamefont{M.}~\bibnamefont{Korkusinski}},
	\bibinfo{author}{\bibfnamefont{P.}~\bibnamefont{Hawrylak}},
	\bibinfo{author}{\bibfnamefont{H.~W.} \bibnamefont{Liu}}, \bibnamefont{and}
	\bibinfo{author}{\bibfnamefont{Y.}~\bibnamefont{Hirayama}},
	\bibinfo{journal}{Scientific Reports} \textbf{\bibinfo{volume}{7}},
	\bibinfo{pages}{43553EP} (\bibinfo{year}{2017}).
	
	\bibitem[{\citenamefont{Yusa et~al.}(2005)\citenamefont{Yusa, Muraki,
			Takashina, Hashimoto, and Hirayama}}]{Yusa05}
	\bibinfo{author}{\bibfnamefont{G.}~\bibnamefont{Yusa}},
	\bibinfo{author}{\bibfnamefont{K.}~\bibnamefont{Muraki}},
	\bibinfo{author}{\bibfnamefont{K.}~\bibnamefont{Takashina}},
	\bibinfo{author}{\bibfnamefont{K.}~\bibnamefont{Hashimoto}},
	\bibnamefont{and} \bibinfo{author}{\bibfnamefont{Y.}~\bibnamefont{Hirayama}},
	\bibinfo{journal}{Nature} \textbf{\bibinfo{volume}{434}},
	\bibinfo{pages}{1001} (\bibinfo{year}{2005}).
	
	\bibitem[{\citenamefont{Hennel et~al.}(2016)\citenamefont{Hennel, Braem, Baer,
			Tiemann, Sohi, Wehrli, Hofmann, Reichl, Wegscheider, R\"ossler
			et~al.}}]{Hennel16}
	\bibinfo{author}{\bibfnamefont{S.}~\bibnamefont{Hennel}},
	\bibinfo{author}{\bibfnamefont{B.~A.} \bibnamefont{Braem}},
	\bibinfo{author}{\bibfnamefont{S.}~\bibnamefont{Baer}},
	\bibinfo{author}{\bibfnamefont{L.}~\bibnamefont{Tiemann}},
	\bibinfo{author}{\bibfnamefont{P.}~\bibnamefont{Sohi}},
	\bibinfo{author}{\bibfnamefont{D.}~\bibnamefont{Wehrli}},
	\bibinfo{author}{\bibfnamefont{A.}~\bibnamefont{Hofmann}},
	\bibinfo{author}{\bibfnamefont{C.}~\bibnamefont{Reichl}},
	\bibinfo{author}{\bibfnamefont{W.}~\bibnamefont{Wegscheider}},
	\bibinfo{author}{\bibfnamefont{C.}~\bibnamefont{R\"ossler}},
	\bibnamefont{et~al.}, \bibinfo{journal}{Phys. Rev. Lett.}
	\textbf{\bibinfo{volume}{116}}, \bibinfo{pages}{136804}
	(\bibinfo{year}{2016}).
	
	\bibitem[{\citenamefont{Watanabe et~al.}(2010)\citenamefont{Watanabe,
			Igarashia, Hashimoto, Kumada, and Hirayama}}]{Watanabe10}
	\bibinfo{author}{\bibfnamefont{S.}~\bibnamefont{Watanabe}},
	\bibinfo{author}{\bibfnamefont{G.}~\bibnamefont{Igarashia}},
	\bibinfo{author}{\bibfnamefont{K.}~\bibnamefont{Hashimoto}},
	\bibinfo{author}{\bibfnamefont{N.}~\bibnamefont{Kumada}}, \bibnamefont{and}
	\bibinfo{author}{\bibfnamefont{Y.}~\bibnamefont{Hirayama}},
	\bibinfo{journal}{Physica E} \textbf{\bibinfo{volume}{42}},
	\bibinfo{pages}{999} (\bibinfo{year}{2010}).
	
	\bibitem[{\citenamefont{{Watanabe} et~al.}(2012)\citenamefont{{Watanabe},
			{Igarashi}, {Kumada}, and {Hirayama}}}]{Watanabe12}
	\bibinfo{author}{\bibfnamefont{S.}~\bibnamefont{{Watanabe}}},
	\bibinfo{author}{\bibfnamefont{G.}~\bibnamefont{{Igarashi}}},
	\bibinfo{author}{\bibfnamefont{N.}~\bibnamefont{{Kumada}}}, \bibnamefont{and}
	\bibinfo{author}{\bibfnamefont{Y.}~\bibnamefont{{Hirayama}}},
	\bibinfo{journal}{arXiv:1210.6223}  (\bibinfo{year}{2012}).
	
	\bibitem[{\citenamefont{Wu et~al.}(2018)\citenamefont{Wu, Wan, Kazakov, Wang,
			Simion, Liang, West, Baldwin, Pfeiffer, Lyanda-Geller et~al.}}]{Wu18}
	\bibinfo{author}{\bibfnamefont{T.}~\bibnamefont{Wu}},
	\bibinfo{author}{\bibfnamefont{Z.}~\bibnamefont{Wan}},
	\bibinfo{author}{\bibfnamefont{A.}~\bibnamefont{Kazakov}},
	\bibinfo{author}{\bibfnamefont{Y.}~\bibnamefont{Wang}},
	\bibinfo{author}{\bibfnamefont{G.}~\bibnamefont{Simion}},
	\bibinfo{author}{\bibfnamefont{J.}~\bibnamefont{Liang}},
	\bibinfo{author}{\bibfnamefont{K.~W.} \bibnamefont{West}},
	\bibinfo{author}{\bibfnamefont{K.}~\bibnamefont{Baldwin}},
	\bibinfo{author}{\bibfnamefont{L.~N.} \bibnamefont{Pfeiffer}},
	\bibinfo{author}{\bibfnamefont{Y.}~\bibnamefont{Lyanda-Geller}},
	\bibnamefont{et~al.}, \bibinfo{journal}{Phys. Rev. B}
	\textbf{\bibinfo{volume}{97}}, \bibinfo{pages}{245304}
	(\bibinfo{year}{2018}).
	
	\bibitem[{foo({\natexlab{a}})}]{footnote:spin_only}
	\bibinfo{note}{For simplicity we consider spin as the extra (non-LL-index)
		degree of freedom, but similar considerations may be applied, e.g., to states
		of different valleys or layers in a bilayer system~(cf.~Ref.~\cite{Cote10}).}
	
	\bibitem[{\citenamefont{Jungwirth et~al.}(1998)\citenamefont{Jungwirth, Shukla,
			Smr\ifmmode~\check{c}\else \v{c}\fi{}ka, Shayegan, and
			MacDonald}}]{Jungwirth98}
	\bibinfo{author}{\bibfnamefont{T.}~\bibnamefont{Jungwirth}},
	\bibinfo{author}{\bibfnamefont{S.~P.} \bibnamefont{Shukla}},
	\bibinfo{author}{\bibfnamefont{L.}~\bibnamefont{Smr\ifmmode~\check{c}\else
			\v{c}\fi{}ka}}, \bibinfo{author}{\bibfnamefont{M.}~\bibnamefont{Shayegan}},
	\bibnamefont{and} \bibinfo{author}{\bibfnamefont{A.~H.}
		\bibnamefont{MacDonald}}, \bibinfo{journal}{Phys. Rev. Lett.}
	\textbf{\bibinfo{volume}{81}}, \bibinfo{pages}{2328} (\bibinfo{year}{1998}).
	
	\bibitem[{\citenamefont{Jungwirth and
			MacDonald}(2001{\natexlab{b}})}]{Jungwirth2001a}
	\bibinfo{author}{\bibfnamefont{T.}~\bibnamefont{Jungwirth}} \bibnamefont{and}
	\bibinfo{author}{\bibfnamefont{A.~H.} \bibnamefont{MacDonald}},
	\bibinfo{journal}{Phys. Rev. B} \textbf{\bibinfo{volume}{63}},
	\bibinfo{pages}{035305} (\bibinfo{year}{2001}{\natexlab{b}}).
	
	\bibitem[{\citenamefont{Fertig et~al.}(1994)\citenamefont{Fertig, Brey,
			C\^ot\'e, and MacDonald}}]{Fertig94}
	\bibinfo{author}{\bibfnamefont{H.~A.} \bibnamefont{Fertig}},
	\bibinfo{author}{\bibfnamefont{L.}~\bibnamefont{Brey}},
	\bibinfo{author}{\bibfnamefont{R.}~\bibnamefont{C\^ot\'e}}, \bibnamefont{and}
	\bibinfo{author}{\bibfnamefont{A.~H.} \bibnamefont{MacDonald}},
	\bibinfo{journal}{Phys. Rev. B} \textbf{\bibinfo{volume}{50}},
	\bibinfo{pages}{11018} (\bibinfo{year}{1994}).
	
	\bibitem[{\citenamefont{Fertig et~al.}(1997)\citenamefont{Fertig, Brey,
			C\^ot\'e, MacDonald, Karlhede, and Sondhi}}]{Fertig97}
	\bibinfo{author}{\bibfnamefont{H.~A.} \bibnamefont{Fertig}},
	\bibinfo{author}{\bibfnamefont{L.}~\bibnamefont{Brey}},
	\bibinfo{author}{\bibfnamefont{R.}~\bibnamefont{C\^ot\'e}},
	\bibinfo{author}{\bibfnamefont{A.~H.} \bibnamefont{MacDonald}},
	\bibinfo{author}{\bibfnamefont{A.}~\bibnamefont{Karlhede}}, \bibnamefont{and}
	\bibinfo{author}{\bibfnamefont{S.~L.} \bibnamefont{Sondhi}},
	\bibinfo{journal}{Phys. Rev. B} \textbf{\bibinfo{volume}{55}},
	\bibinfo{pages}{10671} (\bibinfo{year}{1997}).
	
	\bibitem[{\citenamefont{MacDonald et~al.}(1997)\citenamefont{MacDonald, Fertig,
			and Brey}}]{MacDonald97}
	\bibinfo{author}{\bibfnamefont{A.~H.} \bibnamefont{MacDonald}},
	\bibinfo{author}{\bibfnamefont{H.~A.} \bibnamefont{Fertig}},
	\bibnamefont{and} \bibinfo{author}{\bibfnamefont{L.}~\bibnamefont{Brey}},
	\bibinfo{journal}{Phys. Rev. Lett.} \textbf{\bibinfo{volume}{76}},
	\bibinfo{pages}{2153} (\bibinfo{year}{1997}).
	
	\bibitem[{\citenamefont{Fal'ko and Iordanskii}(1999)}]{Falko99}
	\bibinfo{author}{\bibfnamefont{V.~I.} \bibnamefont{Fal'ko}} \bibnamefont{and}
	\bibinfo{author}{\bibfnamefont{S.~V.} \bibnamefont{Iordanskii}},
	\bibinfo{journal}{Phys. Rev. Lett.} \textbf{\bibinfo{volume}{82}},
	\bibinfo{pages}{402} (\bibinfo{year}{1999}).
	
	\bibitem[{\citenamefont{Mitra and Girvin}(2003)}]{Mitra03}
	\bibinfo{author}{\bibfnamefont{A.}~\bibnamefont{Mitra}} \bibnamefont{and}
	\bibinfo{author}{\bibfnamefont{S.~M.} \bibnamefont{Girvin}},
	\bibinfo{journal}{Phys. Rev. B} \textbf{\bibinfo{volume}{67}},
	\bibinfo{pages}{245311} (\bibinfo{year}{2003}).
	
	\bibitem[{\citenamefont{Kazakov et~al.}(2017)\citenamefont{Kazakov, Simion,
			Lyanda-Geller, Kolkovsky, Adamus, Karczewski, Wojtowicz, and
			Rokhinson}}]{Kazakov17}
	\bibinfo{author}{\bibfnamefont{A.}~\bibnamefont{Kazakov}},
	\bibinfo{author}{\bibfnamefont{G.}~\bibnamefont{Simion}},
	\bibinfo{author}{\bibfnamefont{Y.}~\bibnamefont{Lyanda-Geller}},
	\bibinfo{author}{\bibfnamefont{V.}~\bibnamefont{Kolkovsky}},
	\bibinfo{author}{\bibfnamefont{Z.}~\bibnamefont{Adamus}},
	\bibinfo{author}{\bibfnamefont{G.}~\bibnamefont{Karczewski}},
	\bibinfo{author}{\bibfnamefont{T.}~\bibnamefont{Wojtowicz}},
	\bibnamefont{and} \bibinfo{author}{\bibfnamefont{L.~P.}
		\bibnamefont{Rokhinson}}, \bibinfo{journal}{Phys. Rev. Lett.}
	\textbf{\bibinfo{volume}{119}}, \bibinfo{pages}{046803}
	(\bibinfo{year}{2017}).
	
	\bibitem[{\citenamefont{Haldane}(2009)}]{Haldane09}
	\bibinfo{author}{\bibfnamefont{F.~D.~M.} \bibnamefont{Haldane}},
	\bibinfo{journal}{arXiv:0906.1854}  (\bibinfo{year}{2009}).
	
	\bibitem[{\citenamefont{Park and Haldane}(2014)}]{Park14}
	\bibinfo{author}{\bibfnamefont{Y.}~\bibnamefont{Park}} \bibnamefont{and}
	\bibinfo{author}{\bibfnamefont{F.~D.~M.} \bibnamefont{Haldane}},
	\bibinfo{journal}{Phys. Rev. B} \textbf{\bibinfo{volume}{90}},
	\bibinfo{pages}{045123} (\bibinfo{year}{2014}).
	
	\bibitem[{\citenamefont{Hashimoto et~al.}(2002)\citenamefont{Hashimoto, Muraki,
			Saku, and Hirayama}}]{Hashimoto02}
	\bibinfo{author}{\bibfnamefont{K.}~\bibnamefont{Hashimoto}},
	\bibinfo{author}{\bibfnamefont{K.}~\bibnamefont{Muraki}},
	\bibinfo{author}{\bibfnamefont{T.}~\bibnamefont{Saku}}, \bibnamefont{and}
	\bibinfo{author}{\bibfnamefont{Y.}~\bibnamefont{Hirayama}},
	\bibinfo{journal}{Phys. Rev. Lett.} \textbf{\bibinfo{volume}{88}},
	\bibinfo{pages}{176601} (\bibinfo{year}{2002}).
	
	\bibitem[{\citenamefont{Yang et~al.}(2017)\citenamefont{Yang, Nagase, Hirayama,
			Mishima, Santos, and Liu}}]{YangK17}
	\bibinfo{author}{\bibfnamefont{K.}~\bibnamefont{Yang}},
	\bibinfo{author}{\bibfnamefont{K.}~\bibnamefont{Nagase}},
	\bibinfo{author}{\bibfnamefont{Y.}~\bibnamefont{Hirayama}},
	\bibinfo{author}{\bibfnamefont{T.~D.} \bibnamefont{Mishima}},
	\bibinfo{author}{\bibfnamefont{M.~B.} \bibnamefont{Santos}},
	\bibnamefont{and} \bibinfo{author}{\bibfnamefont{H.}~\bibnamefont{Liu}},
	\bibinfo{journal}{Nature Communications} \textbf{\bibinfo{volume}{8}},
	\bibinfo{pages}{15084} (\bibinfo{year}{2017}).
	
	\bibitem[{\citenamefont{Miyamoto et~al.}(2016)\citenamefont{Miyamoto, Miura,
			Watanabe, Nagase, and Hirayama}}]{Miyamoto16}
	\bibinfo{author}{\bibfnamefont{S.}~\bibnamefont{Miyamoto}},
	\bibinfo{author}{\bibfnamefont{T.}~\bibnamefont{Miura}},
	\bibinfo{author}{\bibfnamefont{S.}~\bibnamefont{Watanabe}},
	\bibinfo{author}{\bibfnamefont{K.}~\bibnamefont{Nagase}}, \bibnamefont{and}
	\bibinfo{author}{\bibfnamefont{Y.}~\bibnamefont{Hirayama}},
	\bibinfo{journal}{Nano Letters} \textbf{\bibinfo{volume}{16}},
	\bibinfo{pages}{1596} (\bibinfo{year}{2016}).
	
	\bibitem[{\citenamefont{Moon et~al.}(1995)\citenamefont{Moon, Mori, Yang,
			Girvin, MacDonald, Zheng, Yoshioka, and Zhang}}]{Moon95}
	\bibinfo{author}{\bibfnamefont{K.}~\bibnamefont{Moon}},
	\bibinfo{author}{\bibfnamefont{H.}~\bibnamefont{Mori}},
	\bibinfo{author}{\bibfnamefont{K.}~\bibnamefont{Yang}},
	\bibinfo{author}{\bibfnamefont{S.~M.} \bibnamefont{Girvin}},
	\bibinfo{author}{\bibfnamefont{A.~H.} \bibnamefont{MacDonald}},
	\bibinfo{author}{\bibfnamefont{L.}~\bibnamefont{Zheng}},
	\bibinfo{author}{\bibfnamefont{D.}~\bibnamefont{Yoshioka}}, \bibnamefont{and}
	\bibinfo{author}{\bibfnamefont{S.-C.} \bibnamefont{Zhang}},
	\bibinfo{journal}{Phys. Rev. B} \textbf{\bibinfo{volume}{95}},
	\bibinfo{pages}{5138} (\bibinfo{year}{1995}).
	
	\bibitem[{foo({\natexlab{b}})}]{footnote:rotations}
	\bibinfo{note}{This is due to the projection of the pseudospin operators to the
		subspace $\{ n \uparrow, m \downarrow\}$ which introduces wave-function
		overlap integrals in the elements of the $\overline{\boldsymbol\tau}$. This
		causes them to be no longer properly normalized generators of pseudospin
		rotations. However, in the following analytic calculations we will always
		assume the limit of smooth pseudospin textures where these deviations can be
		neglected to leading order.}
	
	\bibitem[{foo({\natexlab{c}})}]{footnote:uniform_density}
	\bibinfo{note}{Note that, although the individual LL wave functions are
		products of Gaussians and Hermite polynomials, they are spaced just right to
		produce a system with uniform density when fully filled in a given LL.}
	
	\bibitem[{foo({\natexlab{d}})}]{footnote:HallViscosity}
	\bibinfo{note}{The difference in Hall viscosities of two integer quantum Hall
		fluids is proportional to the difference in their Landau level
		indices~\cite{Avron95,Read09}.}
	
	\bibitem[{foo({\natexlab{e}})}]{footnote:no_minimization}
	\bibinfo{note}{Note that our aim is to expose the generic features of the
		charge and spin densities around domain walls; therefore we do not explicitly
		perform energy minimization.}
	
	\bibitem[{\citenamefont{Avron et~al.}(1995)\citenamefont{Avron, Seiler, and
			Zograf}}]{Avron95}
	\bibinfo{author}{\bibfnamefont{J.~E.} \bibnamefont{Avron}},
	\bibinfo{author}{\bibfnamefont{R.}~\bibnamefont{Seiler}}, \bibnamefont{and}
	\bibinfo{author}{\bibfnamefont{P.~G.} \bibnamefont{Zograf}},
	\bibinfo{journal}{Phys. Rev. Lett.} \textbf{\bibinfo{volume}{75}},
	\bibinfo{pages}{697} (\bibinfo{year}{1995}).
	
	\bibitem[{\citenamefont{Read}(2009)}]{Read09}
	\bibinfo{author}{\bibfnamefont{N.}~\bibnamefont{Read}}, \bibinfo{journal}{Phys.
		Rev. B} \textbf{\bibinfo{volume}{79}}, \bibinfo{pages}{045308}
	(\bibinfo{year}{2009}).
	
\end{thebibliography}
\end{document}